\documentclass[aps,pra,reprint,superscriptaddress]{revtex4-1}
\usepackage{amsmath}
\usepackage{graphicx}
\usepackage{amssymb}
\usepackage{slashed}
\usepackage{dcolumn}
\usepackage{mathrsfs}
\usepackage{bm}
\usepackage{color}
\usepackage{lineno,hyperref}

\begin{document}

\title{Band topology of pseudo-Hermitian phases through tensor Berry connections and quantum metric}
\author{Yan-Qing Zhu}
\affiliation{Guangdong Provincial Key Laboratory of Quantum Engineering and Quantum Materials,
School of Physics and Telecommunication Engineering,
South China Normal University, Guangzhou 510006, China}
\affiliation{Guangdong-Hong Kong Joint Laboratory of Quantum Matter,
Frontier Research Institute for Physics, South China Normal University, Guangzhou 510006, China}
\affiliation{The University of Hong Kong Shenzhen Institute of Research and Innovation, Shenzhen 518057, China}
%\affiliation{Department of Physics and Center of Theoretical and Computational Physics, The University of Hong Kong, Pokfulam Road, Hong Kong, China}

\author{Wen Zheng}
\affiliation{National Laboratory of Solid
State Microstructures, School of Physics, Nanjing University,
Nanjing 210093, China}

\author{Shi-Liang Zhu}
\email{slzhu@nju.edu.cn}
\affiliation{Guangdong Provincial Key Laboratory of Quantum Engineering and Quantum Materials,
School of Physics and Telecommunication Engineering,
South China Normal University, Guangzhou 510006, China}
\affiliation{Guangdong-Hong Kong Joint Laboratory of Quantum Matter,
Frontier Research Institute for Physics, South China Normal University, Guangzhou 510006, China}

\author{Giandomenico Palumbo}
\email{giandomenico.palumbo@gmail.com }
\affiliation{School of Theoretical Physics, Dublin Institute for Advanced Studies,
10 Burlington Road, Dublin 4, Ireland}

\date{\today}

\begin{abstract}
\noindent
Among non-Hermitian systems, pseudo-Hermitian phases represent a special class of physical models characterized by real energy spectra and by the absence of non-Hermitian skin effects.
Here, we show that several pseudo-Hermitian phases in two and three dimensions can be built by employing $q$-deformed matrices, which are related to the representation of deformed algebras. Through this algebraic approach we present and study the pseudo-Hermitian version of well known Hermitian topological phases, raging from two-dimensional Chern insulators and time-reversal-invariant topological insulators to three-dimensional Weyl semimetals and chiral topological insulators.
We analyze their topological bulk states through non-Hermitian generalizations of Abelian and non-Abelian tensor Berry connections and quantum metric. Although our pseudo-Hermitian models and their Hermitian counterparts share the same topological invariants, their band geometries are different. We indeed show that some of our pseudo-Hermitian phases
naturally support nearly-flat topological bands, opening the route to the study of pseudo-Hermitian strongly-interacting systems.
Finally, we provide an experimental protocol to realize our models and measure the full non-Hermitian quantum geometric tensor in synthetic matter.
\end{abstract}
%\pacs{42.50.Pq, 37.30.+i, 03.67.Bg, 76.30.Mi}
\maketitle

\section{Introduction} Non-Hermitian (NH) topological matter is nowadays a very active and rich research field in both theoretical and experimental physics \cite{Esaki2011,Kawabata2019,Ueda2018,Kunst2021,Torres2018,HZhou2019,Fu2018,Zeuner2015,Yoshida2019,Yamamoto2019}.
Among the main physical features of NH topological systems, we remind here
the NH skin effect \cite{Okuma2020,Lee2019,Longhi2019,YYi2020,KZhang2020,Chen,Hughes,Fang}, the lacking of the standard  bulk-edge correspondence \cite{SYao2018b,Kunst2018,SYao2018,HWang2019,Borgnia2020,Imura2019} and the existence of the exceptional
points (EPs), which are related to NH phase transitions \cite{Budich2019,Kawabata2019b,Yoshida2019b,LZhou2018,Nori,Arouca2020,Zhang,Malpuech2}. Such EPs have been recently shown to also exist in quantum systems on curved space \cite{Para2021}.

Generally, these NH systems support complex spectra. However, a special class of NH systems, named pseudo-Hermitian (pH) phases support real spectra \cite{Mostafazadeh2002,Mostafazadeh2003,Kawabata2019,Arouca2020,Das,Ueda,Ohashi,Chong}. This is due to the fact that their Hamiltonians are related to their corresponding complex conjugate through an Hermitian matrix $\eta$ such that $H= \eta H^{\dagger} \eta^{-1}$. In general it is not trivial to identify such matrix. This task becomes easier to achieve if the given pH phase is characterized by suitable symmetries. This is the case of $\mathcal{PT}$ (parity and time reversal)-symmetric systems \cite{Bender1998,Rui2019,Ganainy2018,Leykam,Thomale,Fukui}. However, the $\mathcal{PT}$ symmetry is, differently from the general pH condition, a sufficient but not necessarily condition for the reality of the spectrum \cite{Mostafazadeh2002,Mostafazadeh2003,Brody2016}. Moreover, pH phases do not support any skin effect and are for this reason on the border between standard Hermitian and NH phases.
The band topology of NH topological phases has been intensively investigated by employing several theoretical approaches, for instance, through NH (vector) Berry connections and curvatures \cite{Fu2018,SYao2018,Takane2021,Lieu,Ezawa,YXu2017}.

In Hermitian phases, tensor Berry connections have been recently proposed in Refs~\cite{Palumbo2018,Palumbo2019} as new gauge structures to investigate topological phases. These gauge connections behave as antisymmetric tensor gauge fields in momentum space.
Although they have been employed to characterise several Abelian~\cite{Palumbo2018,Palumbo2019,YQZhu2020,XTan2021,MChen2020,HTDing2020} and non-Abelian topological Hermitian models \cite{Palumbo2021},
their relevance in NH topological phases remains an open question.

The goal of our work is twofold. Firstly, we will build pH versions of several well-known topological phases in two and three dimensions, such as 2D Chern insulators, 2D time-reversal-invariant topological insulators \cite{XLQi2006,Hasan2010,XLQi2011}, 3D Weyl semimetals \cite{XWan2011} and 3D chiral topological insulators \cite{Neupert2012,STWang2014} by employing $q$-deformed matrices \cite{Blohmann2003, Zumino, Steinacker}. These types of matrices are related to the representation of deformed algebras, also known in mathematical physics as quantum groups \cite{Majid,Pflaum}. They play an important role in non-commutative geometry \cite{Majid1,Majid2,Hussin}, supersymmetric quantum field theories \cite{Szabo,Rastelli}, quantum integrable systems \cite{Bogoliubov,Baseilhac}, and more recently they have been considered in SPT spin chains \cite{Quella1,Quella2} and quantum scars \cite{Khemani}. Thus, although we will focus on some specific models, our way to generate pH systems is very general and can be naturally extended to any dimension and in principle to any number of (degenerate or non-degenerate) bands.
Secondly, we will study the band topology of these models through NH generalisations of Abelian and non-Abelian tensor Berry connections and quantum metric. The Abelian version of the latter has been recently proposed in Ref. \cite{DJZhang2019} as a natural extension of the Hermitian one \cite{Provost1980,Grigorenko1992,Zanardi,Gritsev,Malpuech,Peotta,Palumbo,Ozawa,Hsiao,Resta} and its role in band geometry and superfluidity has been studied in Ref.~\cite{PHe2021}. We will show its importance in the band topology of pH phases by generalizing some of the results previously presented in the Hermitian cases \cite{Palumbo2018,Ozawa2018,Roy,Piechon,Lee, Salerno, deJuan,Yang,WChen,Piechon2,Mera-Ozawa1, Mera-Ozawa2,Mera2021}.
Importantly, we also show that the NH skin effects will appear after breaking the pseudo-Hermiticity in the original pH models.
%tensor Berry connections also apply to 3D NH phases with sublattice symmetry that support complex bulk spectra and NH skin effect.
Finally, we will propose an experimental protocol for realizing the pH models in the dilated Hermitian Hamiltonian and detecting the full NH quantum geometric tensor (QGT) that will represent a central ingredient to experimentally detect in synthetic-matter setups the pH topological phases presented in our work.
Our results provide the first theoretical evidence of the deep role of NH tensor Berry connections and quantum metric in the topological characterisation of pH models.\\

%%%%%%%%%%%%%%%%%%%%%%%%%%%%%%%%%%%%%%%%%%%%%
\begin{figure}[htbp]\centering
	\includegraphics[width=8.8cm]{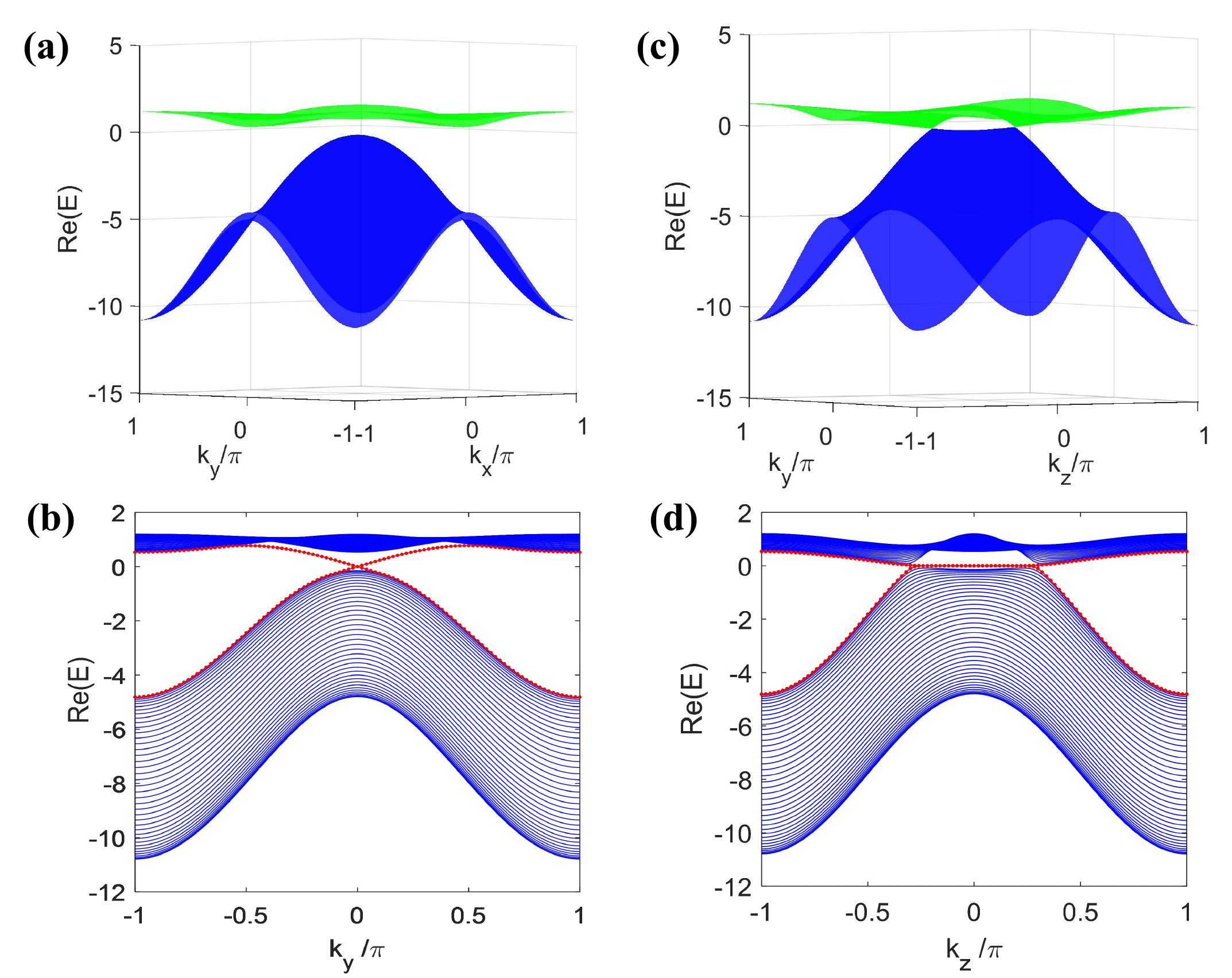}
	\caption{ Energy spectra of the pH Chern/Weyl phase. (a) Bulk spectrum of a 2D pH Chern insulator; (b) The corresponding edge spectrum under open boundary condition along $x$ direction with $M=1.6$, where the chiral edge states (red) emerge. (c) Bulk spectrum of a 3D pH Weyl semimetal on $k_x=0$ plane; (d) The corresponding surface spectrum on $k_y=0$ plane under open boundary condition along $x$ direction with $M=2.6$, where the Fermi-arc surface states (red) are connecting two pH Weyl points. In (a-d), we take the lattice length $L_x=40$ with $v_{x,y,z}=1$ and $q=3$. } \label{pHChern}
\end{figure}
%%%%%%%%%%%%%%%%%%%%%%%%%%%%%%%%%%%%%%%%%%%%%%%

\section{Pseudo-Hermitian Chern insulators and Weyl semimetals}\label{PHChern}
Our general approach to build pH phases from the standard Hermitian ones relies on the concept of deformed algebras and the corresponding $q$-deformed matrices.
Given a generic Hermitian Hamiltonian $H_h$ that carries some global symmetries encoded in an algebra $G$ with generators $T^i$, one can define a corresponding q-deformed algebra $G_q$, with generators $T^i_q$, where $q$ is known as deforming parameter (for $q=1$, one recovers the original algebra $G$). For our purpose, in the examples that we will illustrate below, we will mainly focus on the Clifford algebra by replacing its generators (or part of them depending on the representation and topological class) with the $q$-deformed ones.
Here, we start by considering a two-band Chern insulator whose Hamiltonian is given by
\begin{equation}
\begin{aligned}\label{pHCTIHam}
\tilde{\mathcal{H}}_{CI}(\boldsymbol k)&=d_x\tilde{\sigma}_x+d_y\tilde{\sigma}_y+ d_z\tilde{\sigma}_z,
\end{aligned}
\end{equation}
with the three-component Bloch vector as
\begin{equation}
\begin{aligned}\label{BlochVec}
d_{\mu}=v_{\mu}\sin k_{\mu},
d_z=v_z(M-\sum_{\mu}\cos k_{\mu}),
\end{aligned}
\end{equation}
where $\mu=x,y$, the Fermi velocity $v_{x,y,z}=2J$, $J$ is the hopping amplitude, $M$ is a tunable parameter.  Unlike the standard Pauli matrices in the SU(2) representation, $\tilde{\sigma}_i$ here are the $q$-deformed Pauli matrices which satisfy the $q$-deformed Clifford algebra and take the form \cite{Blohmann2003}
\begin{equation}
\begin{aligned}
\tilde{\sigma}_x&=\left(
                   \begin{array}{cc}
                     0 & a \\
                     b & 0 \\
                   \end{array}
                 \right),
\tilde{\sigma}_y&=\left(
                   \begin{array}{cc}
                     0 & -ia \\
                     ib & 0 \\
                   \end{array}
                 \right),~
\tilde{\sigma}_z=\left(
                   \begin{array}{cc}
                     q^{-1} & 0 \\
                     0 & -q \\
                   \end{array}
                 \right),
 \end{aligned}
\end{equation}
with $a=\sqrt{(1+q^2)/2}$ and $b=\sqrt{(1+q^{-2})/2}$.
When $q\neq1$, these matrices are NH and thus the model in Eq.~\eqref{pHCTIHam} describes a NH Chern insulator ($q=0$ is ill-defined). This model always respects pseudo-Hermiticity ($q>0$), given by
\begin{equation}
\eta_1\tilde{\mathcal{H}}_{CI}^{\dagger}\eta_1^{-1}=\tilde{\mathcal{H}}_{CI},~~\eta_1=\text{diag}(q^{\frac{1}{2}},q^{-\frac{1}{2}}),
\end{equation}
with $\eta_1^{\dagger}=\eta_1$, and $(\eta_1^{-1})^{\dagger}=\eta_1^{-1}$.
Its spectrum is always real and reads
\begin{equation}\label{CISpec}
E(\boldsymbol k)= d d_z\pm\sqrt{ab(d_x^2+d_y^2)+c^2d_z^2},
\end{equation}
where $c=(1+q^2)/(2q)$, and $d=(1-q^2)/(2q)$. Notice that model in Eq.~\eqref{pHCTIHam} corresponds to symmetry class A with pseudo-Hermiticity, which hosts two independent Chern numbers ($\mathbb{Z}\oplus\mathbb{Z}$) in the presence of a real gap in 2D \cite{Kawabata2019}. One of them is defined for $\tilde{\mathcal{H}}_{CI}$ and the other one for the Hermitian model $\eta_1^{-1}\tilde{\mathcal{H}}_{CI}$.
The (first) Chern number of a NH system can be defined through the bi-orthogonal basis as follows\cite{Fu2018},
\begin{equation}
C_n=\frac{1}{2\pi}\int_{\mathbb{T}^2} dk_{\mu}dk_{\nu} F^n_{\mu\nu}
\end{equation}
where the associated NH Berry curvature for the $n$-th Bloch band is given by
\begin{equation}
F^n_{\mu\nu}=\partial_{\mu}A^n_{\nu}-\partial_{\nu}A^n_{\mu},~~A^{n}_{\mu}=\langle u^L_n|i\partial_{\mu}|u^R_n\rangle.
\end{equation}
where $|u^R_n\rangle$ and the $|u^L_n\rangle$ are the Bloch eigenvectors associated to $\tilde{\mathcal{H}}_{CI}$  and $\tilde{\mathcal{H}}^{\dagger}_{CI}$, respectively. For $|M|<2$ ($|M|>2$),  $C_1=-\text{sgn}(M) (C_1=0)$ for the lower band. The topology of the corresponding Hermitian counterpart $\eta_1^{-1}\tilde{\mathcal{H}}_{CI}$ is the same as original one.
The boundary states of this pH phase are given by a propagating chiral model similarly to the Hermitian Chern insulator. This can be shown also analytically by employing a NH version of the Jackiw-Rebbi approach; see Appendix \ref{JBapproch}.
Through the same $q$-deformed Pauli matrices we now study a 3D pH Weyl semimetal, which can be built from $\tilde{\mathcal{H}}_{CI}$ by adding an external term $-v_z\cos k_z$ into $d_z$. A pair of pH Weyl points are located along $k_z$ axis when $1<M<3$. For instance, there is a pair of pH Weyl points at ${\boldsymbol k}_W^{\pm}=(0,0,\pm\frac{\pi}{2})$ when $M=2$. The effective Hamiltonian near these points is given by
\begin{equation}
\tilde{\mathcal{H}}_{W}^{\pm}=v_xq_x^{\pm}\tilde{\sigma}_x+v_yq_y^{\pm}\tilde{\sigma}_y+ v_zq_z^{\pm}\tilde{\sigma}_z,
\end{equation}
where ${\boldsymbol q}_\pm={\boldsymbol k}-{\boldsymbol k}^{\pm}_W$. Each pH Weyl point carries the monopole charge $C_1=+1$ and $C_1=-1$ which are obtained by calculating the Chern number on the 2D sphere enclosing the monopole, respectively. Although our Hermitian and pH two-band models share the same band topology, their band geometry is crucially different. In fact, $q$ gives rise to nearly-flat topological bands shown in Fig. \ref{pHChern}. The presence of nearly-flat Chern bands has important consequences in the interacting regime and opens the route to study pH strongly-interacting phases in lattice models, which will be analyzed in future work.

\begin{figure}[htbp]\centering
	\includegraphics[width=8.0cm]{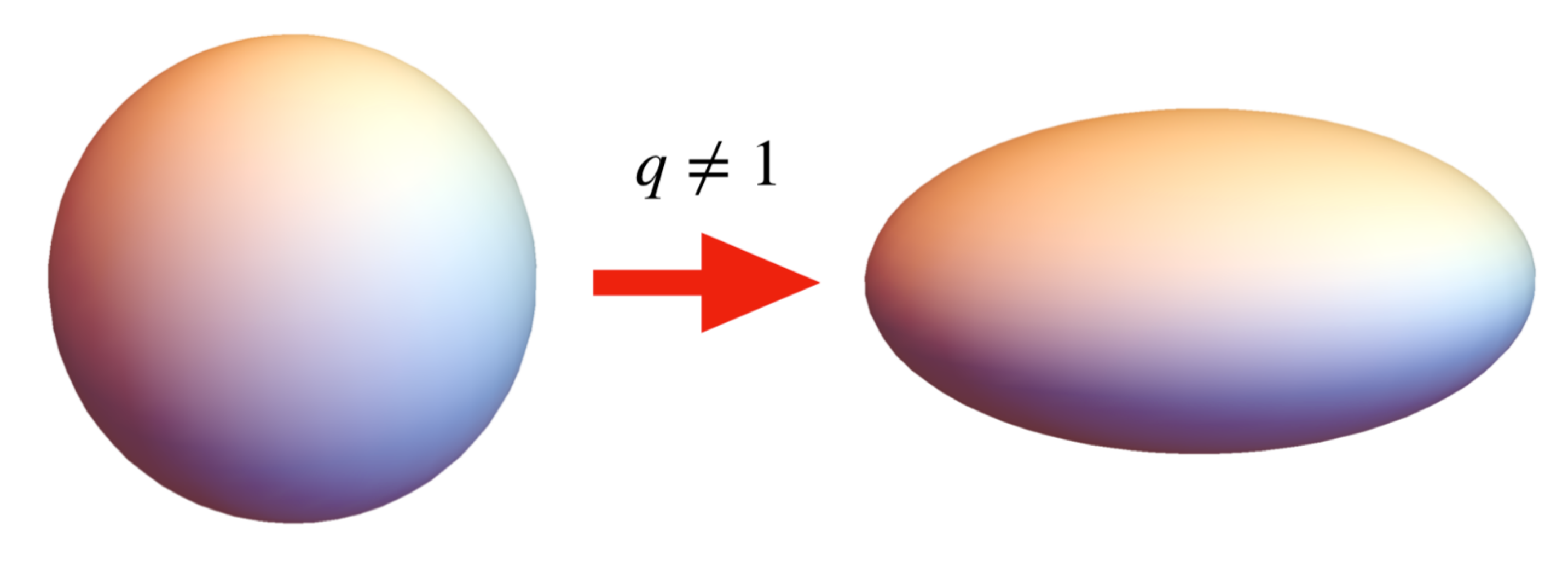}
\caption{The $q$-parameter deforms a 2D sphere (left) to an ellipsoid (right), which are the underlying quantum state manifold identified by the quantum metric associated to Hermitian ($q=1$) and pH ($q\ne 1$) Weyl cones, respectively.}\label{Current}
\end{figure}

Besides Berry connections, another important tool to study the band topology and geometry is given by the quantum metric $g_{\mu\nu}$. In the NH case, $g_{\mu\nu}$ is the real part of the NH QGT $Q_{\mu\nu}^n$ which measures the distance between two nearby density matrices \cite{DJZhang2019} with  the definition,
\begin{equation}
\begin{aligned}
Q^n_{\mu\nu}=\frac{1}{2}[\langle \partial_{\mu} u_n^L|(1-P_n)|\partial_{\nu} u_n^R\rangle+ \langle\partial_{\mu} u_n^R|(1-P_n^{\dagger})|\partial_{\nu} u_n^L\rangle],
\end{aligned}
\end{equation}
where the projection operator $P_n=|u^n_R\rangle\langle u^n_L|$.
Thus, the quantum metric tensor $g_{\mu\nu}=\text{Re}(Q_{\mu\nu}^n)$ while the imaginary part corresponds to the NH Berry curvature, i.e., $F^n_{\mu\nu}=-2\text{Im}(Q^n_{\mu\nu})$.
For our two-band systems there is still a relation between the NH Berry curvature and  the NH quantum metric, i.e.,
\begin{equation}
|F^n_{\mu\nu}|=2\sqrt{g},
\end{equation}
where $g=\det g_{\mu\nu}$ is the determinant of the $2\times2$ NH metric tensor defined for each band in the case of the 2D Chern insulator and in the proper 2D subspace in the case of the 3D Weyl semimetal; see Appendix \ref{ABMetric} . Although the above relation is formally similar to the Hermitian counterpart, in our 3D pH case the NH quantum metric identifies a 2D ellipsoid for $q\neq 1$ differently from the standard 2D sphere in the Hermitian case.
From this point of view, the $q$ deforming parameter has a precise geometric meaning: it deforms the sphere to an ellipsoid by keeping the band topology of the Hermitian model, see Fig.~\ref{Current}.
%%%%%%%%%%%%%%%%%%%%%%%%%%%%%%%%%%%%%%%%%%%%%
%%%%%%%%%%%%%%%%%%%%%%%%%%%%%%%%%%%%%%%%%%%%%%%

\section{Pseudo-Hermitian time-reversal-invariant topological insulators}
 We now construct a pH time-reversal-invariant insulator in 2D. The Hamiltonian is given by
\begin{equation}
\tilde{\mathcal{H}}_{TRI}({\boldsymbol k})=d_x\sigma_z\otimes\tilde{\sigma}_x+d_y\sigma_0\otimes\tilde{\sigma}_y+d_z\sigma_0\otimes\tilde{\sigma}_z,
\end{equation}
where the Bloch vector takes the same form as in Eq.~\eqref{BlochVec}.
The model indeed respects both time-reversal symmetry (TRS) and pseudo-Hermiticity,
\begin{equation}
\hat{T}\tilde{\mathcal{H}}_{TRI}({\boldsymbol k})\hat{T}^{-1}=\tilde{\mathcal{H}}_{TRI}({-\boldsymbol k}),~~\eta_2\tilde{\mathcal{H}}^{\dagger}_{TRI}\eta_2^{-1}=\tilde{\mathcal{H}}_{TRI},
\end{equation}
where $\hat{T}=\sigma_y\otimes\sigma_0\hat{K}$, $\hat{T}^2=-1$, $\eta_2=\sigma_z\otimes \eta_1$, $\{\eta_2,\hat{T}\}=0$. Here $\sigma_i$ are the standard Pauli matrices, $\hat{K}$ is the complex conjugate.  Its doubly degenerate bulk spectrum is similar to Eq.~\eqref{CISpec}. This implies that by tuning the $q$ parameter we can obtain nearly-flat degenerate lower (upper) bands.
Since the system belongs to class AII with pseudo-Hermiticity and its operator anti-commutes with $\hat{T}$, it hosts a $\mathbb{Z}$ invariant \cite{Kawabata2019}. Due to the presence of TRS, the Chern number for $\tilde{\mathcal{H}}_{TRI}$ is zero while the Hermitian counterpart $\eta^{-1}_2\tilde{\mathcal{H}}_{TRI}$ hosts a non-zero Chern number $C_1=-2\text{sgn}(M)$ for $|M|<2$.
However, the topology of the NH topological insulator can not be characterized by the NH Chern number.

Here, we employ the NH version of the non-Abelian tensor Berry connection to characterize this NH $\mathbb{Z}$-class topological phase. In order to build this NH tensor gauge field, we need first to define a proper NH momentum-space Higgs field \cite{Palumbo2021}, which can be built as
\begin{equation}
\Phi= (\tilde{\Phi} \cdot \tilde{\Phi})^{-1/2}\tilde{\Phi},~~
\tilde{\Phi}^{mn}=\langle u_m^L|\eta_2|u_n^R\rangle,
\end{equation}
{where the dot product denotes a matrix product hereafter.}  For the above model we obtain $\Phi=-\sigma_z$.
In this way, the non-Abelian tensor Berry connection for the lower doubly degenerate band is given by
\begin{equation}
\mathbf{B}_{\mu\nu}=\Phi \cdot \mathbf{F}_{\mu\nu},~~\mathbf{F}_{\mu\nu}=\partial_{\mu}\mathbf{A}_{\nu}-\partial_{\mu}\mathbf{A}_{\nu}-i[\mathbf{A}_{\mu},\mathbf{A}_{\nu}],
\end{equation}
where $A^{mn}_{\mu}=\langle u_m^L|i\partial_{\mu}|u_n^R\rangle$ is the associated non-Abelian Berry connection. The corresponding 2D Berry-Zak phase is then given by
\begin{equation}
\begin{aligned}
\Upsilon_{\bf B}=\frac{1}{2\pi}\int_{\mathbb{T}^2}dk_xdk_y~\text{tr}~\mathbf{B}_{xy},
\end{aligned}
\end{equation}
which takes the non-zero value $\Upsilon_{\bf B}=-2\text{sgn}(M)$ when $|M|<2$.
This result is similar to its Hermitian counterpart. { Note that  more possible NH (tensor) Berry connections and quantum metric could be built from the left and right eigenvectors which all give rise to the same topological invariants.}

\begin{figure}[htbp]\centering
	\includegraphics[width=8.8cm]{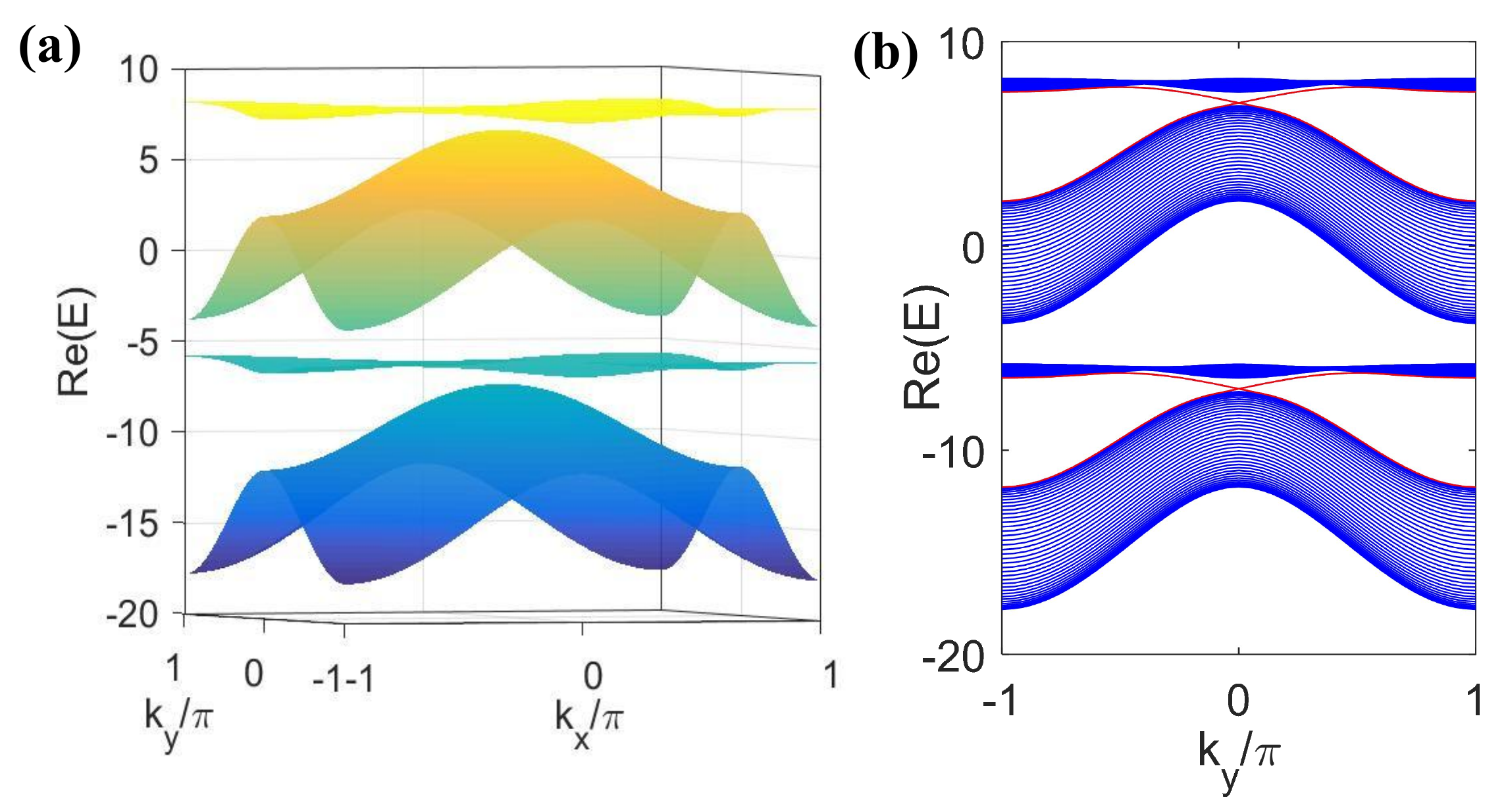}
	\caption{ Energy spectra of the pH Chern phase of $\tilde{\mathcal{H}}'$. (a) Bulk spectrum of a 2D pH Chern insulator; (b) The corresponding edge spectrum under open boundary condition along $x$ direction with $M=1.6$, where the chiral edge states (red) emerge.  We take the lattice length $L_x=40$ with $v_{x,y,z}=1$, $\Delta_0=7$ and $q=3$. } \label{BHZ}
\end{figure}

We now consider a non-Abelian generalization (see, Refs.~\cite{Ma, Neupert} for the Hermitian version) of the NH quantum geometric tensor (QGT) for a $N$-fold degenerate band $Q^{mn}_{\mu\nu}$, {which is defined as
\begin{equation}
\begin{aligned}
Q^{mn}_{\mu\nu}=&\frac{1}{2}\sum_{m,n=1}^N[\langle \partial_{\mu} u_m^L|(1-P)|\partial_{\nu} u_n^R\rangle \\
&+\langle\partial_{\mu} u_m^R|(1-P^{\dagger})|\partial_{\nu} u_n^L\rangle],
\end{aligned}
\end{equation}
where the projection operator is defined as
\begin{equation}
P=\sum_{m=1}^N|u_m^R\rangle\langle u_m^L|,~P^{\dagger}=\sum_{m=1}^N|u_m^L\rangle\langle u_m^R|.
\end{equation}}
From this gauge-invariant quantity we can extract the non-Abelian NH quantum metric $g_{\mu\nu}^{mn}=\text{Re}(Q_{\mu\nu}^{mn})$ and show that the following relation holds
\begin{equation}\label{relation}
|\text{tr}\,\mathbf{B}_{xy}|=2\sqrt{g},
\end{equation}
where now $g=\det (g^{11}_{\mu\nu}+g^{22}_{\mu\nu})$. This represents a further evidence of a deep relation, even at non-Abelian level, between tensor Berry connections and quantum metric; see Appendix \ref{NABMetric}.
We can lift the band degeneracy by adding a TR-broken term, e.g., $\Delta=\Delta_0\,\sigma_z\otimes\sigma_0$. The model $\tilde{\mathcal{H}}'=\tilde{\mathcal{H}}_{TRI}+\Delta$ still respects pseudo-Hermiticity but now belongs to symmetry class A. It has the same topological classification as the pH Chern insulator discussed previously.  $\tilde{\mathcal{H}}'$ hosts a non-zero Chern number for each band $C_1=\pm\text{sgn}(M)$ [a large value of $\Delta_0$ entirely lifts the band degeneracy, {as shown in Fig. \ref{BHZ}(a)], while the Chern number for $\eta_2^{-1}\tilde{\mathcal{H}}'$ remains unchanged. Fig. \ref{BHZ}(b) shows the chiral edge modes emerge in the two subgaps while we consider open boundary condition along $x$ direction.}

%%%%%%%%%%%%%%%%%%%%%%%%%%%%%%%%%%%%%%%%%%%%%
%%%%%%%%%%%%%%%%%%%%%%%%%%%%%%%%%%%%%%%%%%%%%%%

\section{Pseudo-Hermitian chiral topological insulators} We now build a pH 3D insulator that respects both the chiral ($\Gamma$) and sublattice ($S$) symmetries. Its Hamiltonian takes the form,
\begin{eqnarray}\label{cTIHam}
\tilde{\mathcal{H}}_{cTI}({\boldsymbol k})= \lambda_4 d_1 + \lambda_5 d_2 + \tilde{\lambda}_6 d_3+\tilde{\lambda}_7 d_4= \\ \nonumber
\left(
                  \begin{array}{ccc}
                    0 & 0 & d_1-id_2 \\
                    0 & 0 & a(d_3-id_4) \\
                  d_1+id_2 & b(d_3+id_4) & 0 \\
                  \end{array}
                \right),
\end{eqnarray}
where $\lambda_i$ are the Gell-Mann matrices of SU(3) and similarly to the previous cases, here we have replaced $\lambda_6$ and $\lambda_7$ with the q-deformed ones $\tilde{\lambda}_6$ and $\tilde{\lambda}_7$, respectively. Here, $a=\sqrt{(1+q^2)/2}$, $b=\sqrt{(1+q^{-2})/2}$ and the four-component Bloch vector being {
\begin{equation}
\begin{split}
d_1=v_x\sin k_x,d_2=v_y\sin k_y,d_3=v_z\sin(k_z), \\
d_4=v_w(M-\cos k_x-\cos k_y-\cos k_z).
\end{split}
\end{equation} }
For simplicity, we set $v_i=1$ ($i=x,y,z,w$) hereafter.
The system preserves the following symmetries,
\begin{equation}
\begin{aligned}
\eta_3\tilde{\mathcal{H}}^{\dagger}_{cTI}\eta_3^{-1}&=~~\tilde{\mathcal{H}}_{cTI},~\eta_3=\text{diag}(q^{-\frac{1}{2}},q^{\frac{1}{2}},q^{-\frac{1}{2}}),\\
\Gamma\tilde{\mathcal{H}}^{\dagger}_{cTI}\Gamma^{-1}&=-\tilde{\mathcal{H}}_{cTI},
~\Gamma=\text{diag}(q^{-\frac{1}{2}},q^{\frac{1}{2}},-q^{-\frac{1}{2}}),\\
S\tilde{\mathcal{H}}_{cTI}S^{-1}&=-\tilde{\mathcal{H}}_{cTI},~S=\text{diag}(1,1,-1),\\
\end{aligned}
\end{equation}
where the chiral and sublattice operators satisfy the relation: $[\Gamma,S]=0$. The system hosts a real spectrum {
\begin{equation}
E=0,\pm \sqrt{d_1^2+d_2^2+ab(d_3^2+d_4^2)}.
\end{equation} }
This model turns back to the Hermitian case for $q=1$.
It belongs to class AIII with both pseudo-Hermiticity and sublattice symmetry and hosts a $\mathbb{Z}\oplus\mathbb{Z}$ topological invariant.
To characterize its band topology, we introduce a NH Abelian tensor Berry connection, given by
\begin{equation}
B^n_{\mu\nu}=\phi F^n_{\mu\nu},~~\phi=-\frac{i}{2}\log\prod_{\aleph=1}^3 u^R_{n,\aleph},
\end{equation}
where the scalar field $\phi$ is built from $u^R_{n,\aleph}$ which denotes
the components of $|u^R_{n}\rangle$. Then we can define the NH Abelian tensor Berry curvature as follows
\begin{equation}
\mathcal{H}^n_{\mu\nu\lambda}=\partial_{\mu}B^n_{\nu\lambda}
+\partial_{\nu}B^n_{\lambda\mu}+\partial_{\lambda}B^n_{\mu\nu}.
\end{equation}
The associated $\mathcal{DD}$ invariant for the lower band is given by
\begin{equation}\label{DDinv}
\begin{aligned}
\mathcal{DD}=\frac{1}{2\pi^2}\int_{\mathbb{T}^3}dk_xdk_ydk_z~\mathcal{H}_{xyz},
\end{aligned}
\end{equation}
which gives us $\mathcal{DD}=2$ for $|M|<1$, $\mathcal{DD}=-1$ for $1<|M|<3$, and $\mathcal{DD}=0$ elsewhere. Similarly, the Hermitian counterpart $\eta_3^{-1}\tilde{\mathcal{H}}_{cTI}$ has the same topology characterized by the Hermitian $\mathcal{DD}$ invariant.
Even in this case, we can employ the NH quantum metric to identify the band topology of the model.
In fact, in this specific three-band model there exists an important relation between the NH tensor Berry curvature and the NH quantum metric that reads
\begin{equation}
|\mathcal{H}_{xyz}|=4\sqrt{g},
\end{equation}
where $g=\det g_{\mu\nu}$ is the determinant of the $3\times3$ metric tensor defined for each band; see Appendix \ref{ABMetric}.
Moreover, from the same three-band system, we can naturally build a Hamiltonian for a pH tensor monopole. {
Following the model in Eq.~\eqref{cTIHam}, we here analyze the pseudo-Hermitian (pH) 4D tensor semimetal whose model can be obtained by adding an external term $-v_w\cos k_w$ into $d_w$.  A pair of pH tensor monopoles are located along $k_w$ axis when $2<M<4$. For instance, there is a pair of pH tensor monopole at ${\boldsymbol k}_T^{\pm}=(0,0,0,\pm\frac{\pi}{2})$ when $M=3$. The effective Hamiltonian near these points is
\begin{equation}\label{pHTM}
\begin{aligned}
\tilde{\mathcal{H}}^{\pm}_{T}=\left(
                  \begin{array}{ccc}
                    0 & 0 & q_x^{\pm}-iq_y^{\pm} \\
                    0 & 0 & a(q_z^{\pm}-iq_w^{\pm}) \\
                  q_x^{\pm}+iq_y^{\pm} & b(q_z^{\pm}+iq_w^{\pm}) & 0 \\
                  \end{array},
                \right)
\end{aligned}
\end{equation}
where ${\boldsymbol q}_\pm={\boldsymbol k}-{\boldsymbol k}^{\pm}_T$. Each pH tensor monopole carries  topological charge $\mathcal{DD}=+1$ and $\mathcal{DD}=-1$ which are obtained by calculating the $\mathcal{DD}$ on the 3D sphere/ellipsoid enclosing the monopole, respectively.}
However, while in the Hermitian case the metric identifies a 3D sphere, here the NH metric identifies a 3D ellipsoid similarly to the 2D ellipsoid previously discussed.
Notice that the above relation acquires a crucial role in the experimental measurement of the $\mathcal{DD}$ invariant due to the protocol for the experimental measurement of the NH quantum metric that we will present below.

 \section{Skin effects under pseudo-Hermiticity breaking}
{The above models that preserve pseudo-Hermiticity are absence of skin effects since their spectra are purely real.} By breaking the pseudo-Hermiticity, we can explore the NH skin effects in all the previous models.   The topology of these models are characterized by the non-Bloch Hamiltonian $\tilde{\mathcal{H}}(\tilde{\boldsymbol k})$ under different open boundary conditions (OBCs) by replacing the momentum with a complex-momentum as: $\boldsymbol k\rightarrow \tilde{\boldsymbol k}+i\tilde{\boldsymbol k}'$\cite{SYao2018}. The extended bulk-edge correspondence \cite{Imura2019} can be established, and the corresponding phase diagrams are modified under different OBCs. {In this section, we discuss the skin effects in these models. For concreteness and simplicity, we mainly mention on the NH Chern phase below.

Let us add a perturbation $\Delta=i\gamma_x\tilde{\sigma}_x+i\gamma_y\tilde{\sigma}_y+i\gamma_z\tilde{\sigma}_z$ into the 2D pH Chern insulator considered in Sec. \ref{PHChern}, i.e., $\mathcal{H}_{CI}({\boldsymbol k})=\tilde{\mathcal{H}}_{CI}({\boldsymbol k})+\Delta$, which we call $q$-deformed NH Chern insulator. Its spectrum becomes complex and reads,
\begin{equation}
\begin{aligned}
E_{\pm}=d(d_z+i\gamma_z)\pm
\sqrt{\sum_{j=x,y}\left(\epsilon_j+c^2(d_z^2-\gamma_z^2)+2ic^2\gamma_zd_z\right)},
\end{aligned}
\end{equation}
where $\epsilon_j=abd_{j}^2-ab\gamma_j^2+2iab\gamma_jd_j$.
When $\gamma_i=0$, the phase transition is at $M=2$. We below focus on $M$ being close to 2, i.e., $\gamma_{x,y,z}$ sufficiently weak.
The gapped regions ($\text{Re}(E)\neq 0$) of  this model are found to be $M>M_+$ and $M<M_-$, where $M_{\pm}$ have simple expression when $\gamma_z=0$ hereafter,
\begin{equation}
M_{\pm}=2\pm\sqrt{\gamma_x^2+\gamma_y^2}.
\end{equation}
The Bloch phase boundaries are $M=M_{\pm}$, where the gap
closes at ${\boldsymbol k}=(0,0)$.  Since the conventional bulk-edge correspondence is not valid, we have use the non-Bloch theory to study the extended bulk-edge correspondence for the system under OBC.

Inspired by the idea presented in Ref.~\cite{SYao2018}, we can use the non-Bloch theory based on complex-valued wave vectors to investigate its intriguing phase with OBCs. The low-energy continuum model of this NH Chern insulator is given by
\begin{equation}
\begin{aligned}
\mathcal{H}(\boldsymbol k)=(k_x+i\gamma_x)\tilde{\sigma}_x+(k_y+i\gamma_y)\tilde{\sigma}_y\\
+\left(M-2+\frac{k_x^2+k_y^2}{2}\right)\tilde{\sigma}_z.
\end{aligned}
\end{equation}
In  view of this skin effect, we take a complex-valued momentum to describe open-boundary eigenstates: ${\boldsymbol k}\rightarrow \tilde{\boldsymbol k}+i\tilde{{\boldsymbol k}}'$, where the imaginary part $\tilde{\boldsymbol k}'$ takes the simple form $\tilde{k}_j'=-\gamma_j$ for small $\tilde{\boldsymbol k}$ in this model. Firstly, we consider the full OBC along both $x$ and $y$ directions. The corresponding non-Bloch Hamiltonian reads
\begin{equation}
\tilde{\mathcal{H}}(\tilde{\boldsymbol k})=\tilde{k}_x\tilde{\sigma}_x+\tilde{k}_y\tilde{\sigma}_y+\left(\tilde{M}+\frac{\tilde{k}_x^2+\tilde{k}_y^2}{2}-i\sum_{j=x,y}\gamma_j\tilde{k}_j\right)\tilde{\sigma}_z.
\end{equation}
with $\tilde{M}=M-2-(\gamma_x^2+\gamma_y^2)/2$.  When $\gamma_{x,y,z}=\gamma$,  the phase diagram with small $\gamma_j$ can be modified: topological nontrivial region $\tilde{M}<0$ becomes $M<2+\gamma^2$, and the phase boundary: $M_c=2+\gamma^2$, see Fig. \ref{Phasediagram1}(a).
%%%%%%%%%%%%%%%%%%%%%%%%%%%%%%%%%%%%%%%%%%%%%
\begin{figure}[htbp]\centering
\includegraphics[width=8.8cm]{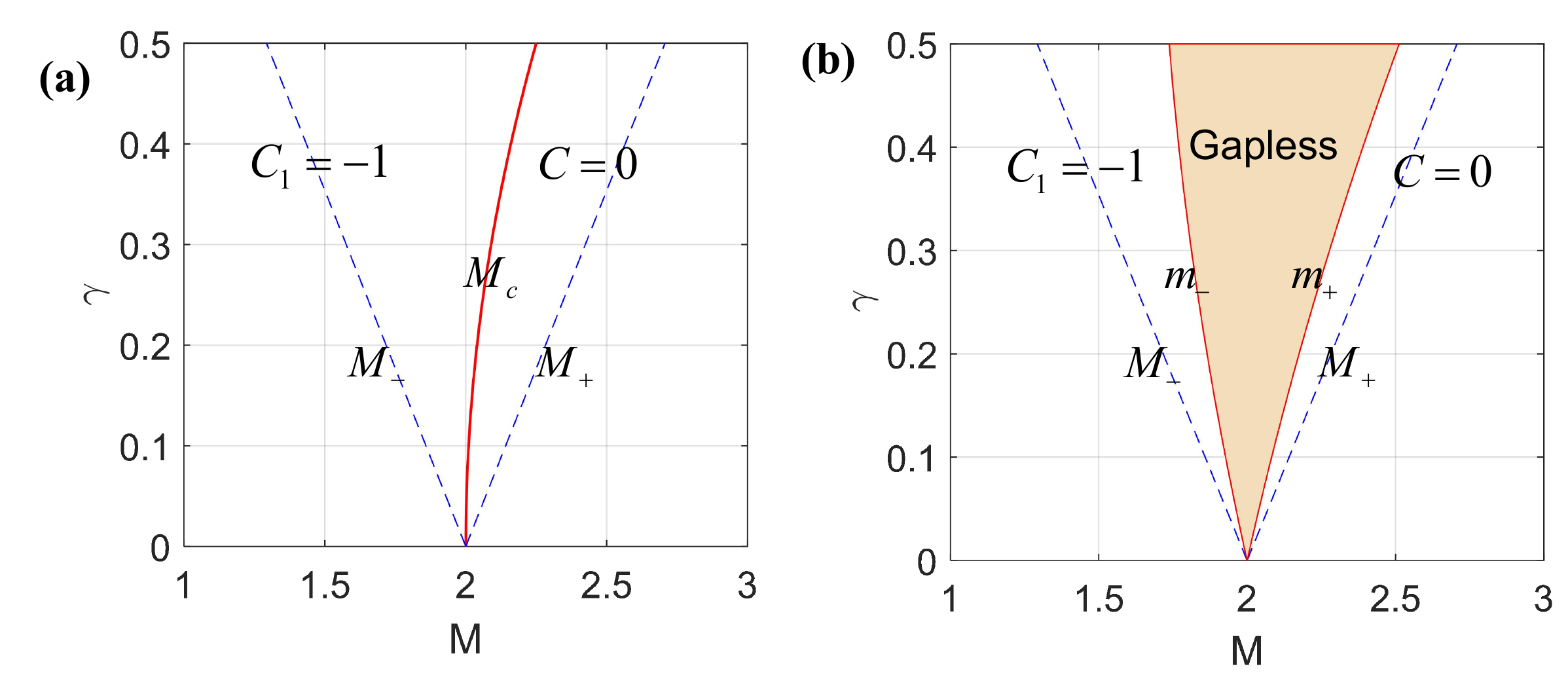}
 \caption{(Color online)  Phase diagram for OBC along (a) all the directions;(b) $x$ direction. (a)The phase boundary marked by red solid line is now modified by $M_c$. (b) The phase boundaries marked by red solid lines are modified by $m_{\pm}$. The dashed lines ($M_{\pm}$) denote the phase boundaries for the Bloch Hamiltonian under periodic boundary condition. In (a-d), we take $q=3$.} \label{Phasediagram1}
\end{figure}
%%%%%%%%%%%%%%%%%%%%%%%%%%%%%%%%%%%%%%%%%%%%%%%
We then consider the OBC just along $x$ direction. Here we just replace $k_{x}$ by $\tilde{k}_x-i\gamma_x$, then we obtain the corresponding non-Bloch Hamiltonian $\tilde{\mathcal{H}}(\tilde{k}_x,k_y)$, given by
\begin{equation}
\begin{aligned}
\tilde{\mathcal{H}}(\tilde{k}_x,k_y)=\tilde{k}_x\tilde{\sigma}_x+(k_y+i\gamma_y)\tilde{\sigma}_y\\
+\left(\tilde{M}+\frac{\tilde{k}_x^2+k_y^2}{2}-i\gamma_x\tilde{k}_x\right)\tilde{\sigma}_z,
\end{aligned}
\end{equation}
with $\tilde{M}=M-2-\gamma_x^2/2$. The phase boundaries are now modified as $m_{\pm}=2+\gamma^2/2\pm\sqrt{ab}/{c}\gamma$ while we take $\gamma_x=\gamma_y=\gamma$.  The topological nontrivial (trivial) region is given by $M<m_-$ ($M>m_+$) while the system is gapless in the region $M\in [m_-,m_+]$, as shown in Fig. \ref{Phasediagram1}(b).
Note that we can further study the edge spectra of this model. Fig. \ref{edgemode_NHChern} shows the numerics for the system under OBC along $x$ direction.  The system hosts a real gap and supports the NH skin modes live in the both left and right ends.
%%%%%%%%%%%%%%%%%%%%%%%%%%%%%%%%%%%%%%%%%%%%%
\begin{figure}[htbp]\centering
\includegraphics[width=8.8cm]{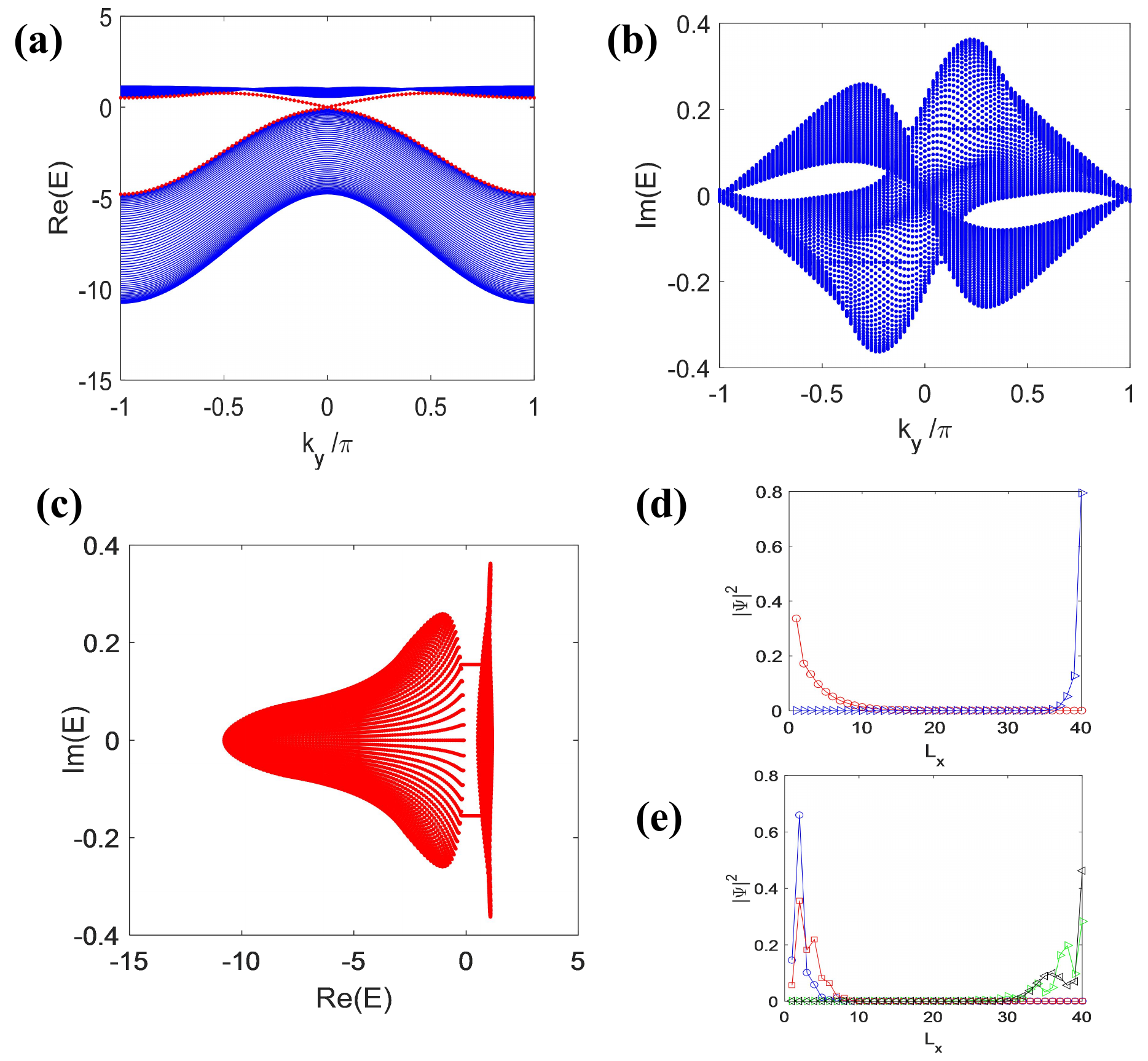}
 \caption{(Color online) Edge spectrum of $\mathcal{H}_{CI}$ under open boundary condition along $x$ direction. (a)Real part of the edge spectrum;(b)Imaginary part of the edge spectrum. (c)The edge spectra hosts a real gap with the edge modes connecting them. The chiral mode colored in red located on each boundary, as shown in (d). (e) Skin modes (colored in black) in the bulk live in  both left and right ends. The parameters we used above are: $M=1.6$, $\gamma=0.2$, $v_{x,y,z}=1$, $L_x=40$.} \label{edgemode_NHChern}
\end{figure}
%%%%%%%%%%%%%%%%%%%%%%%%%%%%%%%%%%%%%%%%%%%%%%%
Similarly, we can explore the Weyl semimetal phase by introducing a same perturbation in the presence of skin effects as shown above. The phase diagram will be modified regarding on different OBCs.

 On the other band, the NH version of the 2D time-reversal-invariant topological insulators and 3D chiral insulators in the presence of skin effect can be also studied through the similar way, i.e., considering the complex Bloch vector  $d_j+i\gamma_j$.  It is noteworthy that a NH sublattice topological insulator will be obtained because the perturbation simultaneously breaks the pseudo-Hermiticity and chiral symmetries. The model now supports the NH skin effect and the phase diagram will be also modified according to different OBCs.
 In Appendix \ref{SLInsulator}, we give another sublattice topological model which hosts the similar topological properties and a rich phase diagram belongs to class A with the sublattice symmetry.}

\section{Experimental proposals}
In general, a NH Hamiltonian can be realized through a dilated Hermitian Hamiltonian \cite{Gunther2007,Gunther2008,YWu2019,YChu2020} with a higher dimensional Hilbert space in the synthetic systems, e.g., nitrogen-vacancy center in diamond \cite{YWu2019,MYu2020}, superconducting circuits \cite{XTan2018,XTan2021}, atomic systems \cite{DWZhang2018,Ueda2018}, etc.  For a pH system $\tilde{\mathcal{H}_s}$, we can realize a dilated Hamiltonian $H_{s,a}$ (system/ancilla qubit $s$/$a$) through the Naimark-Dilation approach, given by
\begin{equation}
H_{s,a}=\sigma^a_0\otimes \Lambda+\sigma^a_y\otimes \Gamma,
\end{equation}
where $\Lambda=\beta^{-1}(\tilde{\mathcal{H}}_s\eta+\eta^{-1}\tilde{\mathcal{H}}_s)$, $\Gamma=i\beta^{-1}(\tilde{\mathcal{H}}_s-\tilde{\mathcal{H}}_s^{\dagger})$, and the constant $\beta=\eta+\eta^{-1}$. {The relevant eigenvector governed by $i\partial_t\Psi(t)=H_{s,a}\Psi(t)$ takes the form,
\begin{equation}\nonumber
\begin{aligned}
\Psi(t)&=\alpha(t)\left(
          \begin{array}{c}
            |\psi(t)\rangle \\
            |\chi(t)\rangle \\
          \end{array}
        \right)\\
&=\alpha(t)\left(|0\rangle_a\otimes|\psi(t)\rangle+|1\rangle_a\otimes|\chi(t)\rangle\right),
\end{aligned}
\end{equation}
where $\alpha(t)$ is the normalization factor,  $|\psi(t)\rangle$ is an effective pH dynamic of the system $s$ which satisfies the following Schr\"odinger equation:
\begin{equation}
i\partial_t|\psi(t)\rangle=\tilde{\mathcal{H}_s}|\psi(t)\rangle,
\end{equation}
while $|\chi(t)\rangle=\eta^{-1}|\psi(t)\rangle$ is nothing but the left vector of $\tilde{\mathcal{H}}^{\dagger}_s$. After doing the post-seclection measurement on the ancilla qubit state $|0\rangle_a$ ($|1\rangle_a$), we could observe and measure the relevant quantities during this effective pH dynamic of
$|\psi(t)\rangle$ ($|\chi(t)\rangle$).} Therefore, a pH Weyl/tensor monopole can be realized in an extended four/six-level Hermitian system.
{Moreover, a generic NH system breaking the pseudo-Hermiticity could also be flexibly realized by a similar but more general way through a dilated Hermitian system which could help to observe the transition from the unbroken to the broken pseudo-Hermiticity; see Appendix \ref{Experiment} for details.}

Inspired by the current experiments and theories on detecting the full Abelian QGT in Hermitian systems \cite{Ozawa2018,MYu2020,XTan2019}, we here demonstrate that we can extract the full QGT in the pH systems through Rabi oscillations; see Appendix \ref{Dection}. By introducing the linear (elliptical) parametric modulations, one could extract the quantum metric tensor (Berry curvature) through the Rabi oscillations. The main different here is that the oscillation between two resonant levels corresponds the bi-orthogonal quantum states. It seems that one could not directly extract the Rabi frequency only from the system $\tilde{\mathcal{H}}_s$.
%Principally, one have to realize both the Hamiltnoians $\tilde{\mathcal{H}}_s$ and  $\tilde{\mathcal{H}}_s^{\dagger}$ for the generic NH systems.
However, thanks to the pseudo-Hermiticity condition, the observable quantity is now determined by
\begin{equation}\label{probability}
|c_{n}(t)|^2=|\langle u_{n}^L(0)|\psi^R(t)\rangle|^2=|\langle u_{n}^R(0)|\eta^{-1}|u_{1}^R(t)\rangle|^2.
\end{equation}
Here, we have employed the relation $|u_n^L\rangle=\eta^{-1}|u_n^R\rangle$ for the pH condition.  Experimentally, we first realize $\tilde{\mathcal{H}}_s$ through the dilated system $H_{s,a}$. After the post-selection measurement, we focus on the dynamic of system $s$, i.e.,  preparing an initial state $|\psi^R(0)\rangle=|u_{1}^R(0)\rangle$ at the lower band and then introducing the parameter modulations.   {For the pH models presented in this paper, $\eta$ is always diagonal. Thus, the probablity is simply expressed as $|c_n(t)|^2=\sum_i(\eta^{-1})^{ii}|n_i(t)|^2$ with $|n_i|^2$ is the population for each basis of $N$-dimensional indentity matrix. However, for a generic pH system, $\eta$ is not always diagonal. Instead, we could extract the probability $|c_n(t)|^2=\text{tr}\left(\rho_n(0)\eta^{-1}\rho_1(t)\eta^{-1}\right)$ with the density matrix $\rho_n(t)=|u^R_n(t)\rangle\langle u^R_n(t)|$ that could be measured from the quantum state tomography technique. Finally, the coherent Rabi frequency can be extracted from the oscillation presented in Eq.~\eqref{probability}.}

\section{Conclusion and outlook}
Summarizing, we have shown that several pH phases can be built by employing $q$-deformed matrices associated to deformed algebras. We have studied the band topology of these models by employing the NH extension of both tensor Berry connections and quantum metric. Our results show the central role of geometric and higher-order gauge fields in the band topology of NH phases.
Moreover, we have provided a measurement protocol for the full NH Abelian QGT that will help to detect pH topological phases in suitable synthetic-matter setups. Notice that other detecting methods \cite{Ozawa2018,XTan2019,MYu2020} for the Abelian QGT in the Hermitian case can also be extended to our pH systems, meanwhile,  the method presented in this paper can be directly generalized to extract the full non-Abelian QGT of general pH systems with degenerate bands.
Several directions will be considered in future work, such as the exploration of novel pH topological superconductors, pH nodal-line and nodal-surface semimetals, and pH higher-order topological phases.

\begin{acknowledgments}
The authors are pleased to acknowledge discussions with Bruno Mera and Peng He.  Work in China is supported by the Key Area Research and Development Program of Guang Dong Province (Grant No. 2019B030330001), the National Natural Science Foundation of China (Grants No. 12074180 and No. U1801661), and the Key Project of Science and Technology of Guangzhou (Grant No. 201804020055).
\end{acknowledgments}

\begin{appendix}
\section{Jackiw-Rebbi approach for the edge states of pseudo-Hermitian Chern insulators}\label{JBapproch}
Here, we consider the continuum version of the 2D pseudo-Hermitian Chern insulator discussed in the main text. Its linearized limit is described by a 2D Dirac-like model
\begin{eqnarray}
H_D=-i (\tilde{\sigma}_x \partial_x+\tilde{\sigma}_y \partial_y) + m(x) \tilde{\sigma}_z,
\end{eqnarray}
where we identify a domain wall at $x=0$ where the spatially-varying mass term $m(x)$ changes sign.
On that straight line, $k_y$ is still a good quantum number. Moreover, we can solve the above model with the following wavefunction $\psi=(u(x), v(x))^{T} e^{i k_y y}$ with energy $E=k_y$. Because we are looking for zero-energy solutions, we pick up $k_y=0$ such that we need to solve the following coupled equations
\begin{eqnarray}
\begin{aligned}
-(m(x)/q) u + \sqrt{(1+q^2)/2}\, \partial_x v=0, \\
-\sqrt{(1+q^{-2})/2}\, \partial_x u + q\, m(x) v=0.
\end{aligned}
\end{eqnarray}
A solution of this system is given by
\begin{eqnarray}
v(x) \propto e^{-(1/A) \int dx\, m(x)},
\end{eqnarray}
with the physical boundary conditions $\lim_{x\rightarrow \pm \infty} v(x)=0$ and $A=\sqrt{(1+q^2)/2q}$.
Thus, the above expression clearly shows that the state is localized at the edge of the system for any positive value of $q$ ($q>0$). This is the pseudo-Hermitian generalization of the Jackiw-Rebbi mechanism \cite{Jackiw} and allows us to analytically check that the bulk-edge correspondence still holds in our system while the NH skin effect is completely absent.

\section{Abelian quantum metric tensor for Dirac/Weyl/Chiral Hamiltonians}\label{ABMetric}
The components of the NH quantum metric tensor (QMT) can be calculated from the NH quantum geometry tensor (QGT), i.e., $g_{\mu\nu}=\frac{1}{2}(Q^n_{\mu\nu}+Q^n_{\nu\mu})$. For simplicity, we consider here the continuum model of a 3D pH Weyl point
\begin{equation}\label{pHWeyl}
\begin{aligned}
\tilde{\mathcal{H}}_W&=k_x\tilde{\sigma}_x+k_y\tilde{\sigma}_y+k_z\tilde{\sigma}_z\\
&=d  k_z\sigma_0+k_x\tilde{\sigma}_x+k_y\tilde{\sigma}_y+c k_z\sigma_z,
\end{aligned}
\end{equation}
with $c=\frac{1+q^2}{2q}$, $d=\frac{1-q^2}{2q}$, $a=\sqrt{(1+q^2)/2}$ and $b=\sqrt{(1+q^{-2})/2}$ ($q>0$). Direct calculation shows that
\begin{equation}
\begin{aligned}\label{S3}
g_{xx}&=~~\frac{ab((ab+k_y^2)+c^2k_z^2)}{4(ab(k_x^2+k_y^2)+c^2k_z^2)^2},\\
g_{yy}&=~~\frac{ab((ab+k_x^2)+c^2k_z^2)}{4(ab(k_x^2+k_y^2)+c^2k_z^2)^2},\\
g_{zz}&=~~\frac{abc^2(k_x^2+k_y^2)}{4(ab(k_x^2+k_y^2)+c^2k_z^2)^2},\\
g_{xy}&=-\frac{a^2b^2k_xk_y}{4(ab(k_x^2+k_y^2)+c^2k_z^2)^2},\\
g_{xz}&=-\frac{abc^2k_xk_z}{4(ab(k_x^2+k_y^2)+c^2k_z^2)^2},\\
g_{yz}&=-\frac{abc^2k_yk_z}{4(ab(k_x^2+k_y^2)+c^2k_z^2)^2}.
\end{aligned}
\end{equation}
The determinants of the  QMT in the proper $2\times 2$ subspace are given by
\begin{equation}
\begin{aligned}
2\sqrt{\det g_{xy}}=\frac{abc|k_z|}{2((ab(k_x^2+k_y^2)+c^2k_z^2))^{3/2}}=|F_{xy}^1|,\\
2\sqrt{\det g_{xz}}=\frac{abc|k_y|}{2((ab(k_x^2+k_y^2)+c^2k_z^2))^{3/2}}=|F_{xz}^1|,\\
2\sqrt{\det g_{yz}}=\frac{abc|k_x|}{2((ab(k_x^2+k_y^2)+c^2k_z^2))^{3/2}}=|F_{yz}^1|,\\
\end{aligned}
\end{equation}
By doing the dimension reduction, i.e, $k_z\rightarrow m$, we obtain a 2D pH massive Dirac cone where only the $\det g_{xy}$ survives. For convenience, especially to calculate the topological charge enclosing by  $\mathbb{S}^2$, we  parameterize the momentum/Bloch vector through the ellipsoidal coordinates
\begin{equation}
\begin{aligned}
k_x=\frac{k}{\sqrt{ab}}\sin\theta\cos\varphi,~ k_y=\frac{k}{\sqrt{ab}}\sin\theta\sin\varphi,~
k_z=\frac{k}{c}\cos\theta,
\end{aligned}
\end{equation}
where $k=\sqrt{ab(k_x^2+k_y^2)+c^2k_z^2}$, then we obtain the corresponding components of QMT which take the same form as the Hermitian case,
\begin{equation}\label{QMT}
g_{\theta\theta}=\frac{1}{4},g_{\varphi\varphi}=\frac{\sin^2\theta}{4},g_{\theta\varphi}=g_{\varphi\theta}=0,
\end{equation}
 with the relation $|F^1_{\theta\varphi}|=2\sqrt{\det g_{\theta\varphi}}=\sin\theta/2$.
Similarly to the 2D/3D Dirac/Weyl case, we can employ the same calculations for the 3D/4D pH chiral/tensor case. The tensor Berry curvature of a pH tensor monopole described by $\tilde{\mathcal{H}}_T^+({\boldsymbol q})$ is given by
\begin{equation}
\begin{aligned}
\mathcal{H}_{xyz}&=~\frac{abq_w}{(q_x^2+q_y^2+ab(q_z^2+q_w^2))^2},\\
\mathcal{H}_{yzw}&=-\frac{abq_x}{(q_x^2+q_y^2+ab(q_z^2+q_w^2))^2},\\
\mathcal{H}_{zwx}&=~\frac{abq_y}{(q_x^2+q_y^2+ab(q_z^2+q_w^2))^2},\\
\mathcal{H}_{wxy}&=-\frac{abq_z}{(q_x^2+q_y^2+ab(q_z^2+q_w^2))^2}.\\
\end{aligned}
\end{equation}
After taking the following parameter transformation,
\begin{equation}
\begin{aligned}\label{PTransform}
&q_x=q\cos\theta_1,q_y=q\sin\theta_1\cos\theta_2,\\
&q_z=\frac{q}{\sqrt{ab}}\sin\theta_1\sin\theta_2\cos\varphi,\\
&q_w=\frac{q}{\sqrt{ab}}\sin\theta_1\sin\theta_2\sin\varphi,
\end{aligned}
\end{equation}
with $q=\sqrt{q_x^2+q_y^2+ab(q_z^2+q_w^2)}$, we obtain the QMT with the same expression like the Hermitian case\cite{XTan2021,MChen2020} which means the relation $|\mathcal{H}_{\theta_1\theta_1\varphi}|=4\sqrt{\det g_{\theta_1\theta_2\varphi}}=\sin^2\theta_1\sin\theta_2$ is still valid.

\section{Relation between the non-abelian quantum metric and the tensor Berry connection}\label{NABMetric}

Without loss of generality, we here consider a 2D massive pH Dirac cone which is the continuum version of Eq.~(10) in the main text and expressed as
\begin{equation}
\tilde{\mathcal{H}}_D({\boldsymbol k})=k_x\sigma_z\otimes\tilde{\sigma}_x+k_y\sigma_0\otimes\tilde{\sigma}_y+m\sigma_0\otimes\tilde{\sigma}_z,
\end{equation}
The components of the QMT for the lower doubly degenerate band are given by
\begin{equation}
\begin{aligned}
g_{xx}^{11}&=g_{xx}^{22}=~~\frac{ab((ab+k_y^2)+c^2m^2)}{4(ab(k_x^2+k_y^2)+c^2k_z^2)^2},\\
g_{yy}^{11}&=g_{yy}^{22}=~~\frac{ab((ab+k_x^2)+c^2m^2)}{4(ab(k_x^2+k_y^2)+c^2k_z^2)^2},\\
g_{xy}^{11}&=g_{xy}^{22}=g_{yx}^{11}=g_{yx}^{22}\\
&=-\frac{a^2b^2k_xk_y}{4(ab(k_x^2+k_y^2)+c^2m^2)^2},\\
&g_{\mu\nu}^{mn}=0, \text{for}~m\neq n,
\end{aligned}
\end{equation}
where the parameters $a$, $b$, $c$ take the same form in Eq.~\eqref{S3}. By taking $g_{\mu\nu}=\sum_{n=1,2}g_{\mu\nu}^{nn}$, we obtain the following relation
\begin{equation}\label{relation}
2\sqrt{\det g_{\mu\nu}}=|\text{tr}\,\mathbf{B}_{xy}|,
\end{equation}
where the non-Abelian tensor Berry  connection $\mathbf{B}_{xy}=-\sigma_z\mathbf{F}_{xy}$ is associated with the non-Abelian Berry curvature $\mathbf{F}_{xy}$,
\begin{equation}
\begin{aligned}
&\mathbf{F}_{xy}=\left(
                  \begin{array}{cc}
                    F_{xy}^{11} & 0 \\
                    0 & F_{xy}^{22} \\
                  \end{array}
                \right),\\
                &F_{xy}^{11}=-F_{xy}^{22}\\
                &=-\frac{abcm}{2(ab(k_x^2 + k_y^2)+c^2m^2)^{3/2}}.
\end{aligned}
\end{equation}
Here, we emphasize that the relation \eqref{relation} between $g_{xy}$ and $\mathbf{B}_{xy}$ always exists even for the lattice model discussed in the main text.

\section{Sublattice topological insulators}\label{SLInsulator}
We provide here an example of 3D NH topological insulator that supports the sublattice symmetry and complex spectrum.
This four-band model represents a 3D NH version of the model in Ref.~\cite{YQZhu2020} and its Hamiltonian is given by
\begin{equation}
\begin{aligned}\label{SIHam}
\mathcal{H}_{SI}(\boldsymbol k)&=(d_1+i\gamma_x)\tilde{\Gamma}_x+(d_2+i\gamma_y)\tilde{\Gamma}_y\\
&+(d_3+i\gamma_z)\tilde{\Gamma}_z+(d_4+i\gamma_w)\tilde{\Gamma}_w,
\end{aligned}
\end{equation}
with the four-component Bloch vector given in Eq.~(18) in the main text. The matrices $\tilde{\Gamma}_j$ are given by
\begin{equation}\nonumber
\begin{aligned}
\tilde{\Gamma}_x=\sigma_0\otimes\sigma_x+a\sigma_x\otimes\sigma_0,~~
\tilde{\Gamma}_y=\sigma_y\otimes\sigma_z+a\sigma_z\otimes\sigma_y,\\
\tilde{\Gamma}_z=\sigma_0\otimes\sigma_y+a\sigma_y\otimes\sigma_0,~~
\tilde{\Gamma}_w=\sigma_x\otimes\sigma_z+a\sigma_z\otimes\sigma_x,\\
 \end{aligned}
\end{equation}
where $\sigma_i$ are the standard Pauli matrices, $a$ is a constant parameter.
For concreteness and convenience, hereafter we consider the situation when $M>1$.
%Notice, these matrices satisfy the Clifford algebra only for $a=0$.
This Hamiltonian always preserves the sublattice symmetry $\{S,\mathcal{H}_{SI}\}=0$ with $S=\sigma_z\otimes\sigma_z$.
Its spectrum reads
\begin{equation}
E(\boldsymbol k)=\pm (1\pm a)\sqrt{\sum_{j=1}^4 (d_j^2-\gamma_j^2+2i\gamma_jd_j)}.
\end{equation}
When $\gamma_i=0$, the model has a topological phase transition point at $M=3$. We below focus on $M$ being close to 3, i.e., $\gamma_{x,y,z,w}$ sufficiently weak.
The gapped regions ($\text{Re}(E)\neq 0$) of  this model are found to be $M>M_+$ and $M<M_-$, where $M_{\pm}$ have simple expression when $\gamma_w=0$,
\begin{equation}
M_{\pm}=3\pm\sqrt{\gamma_x^2+\gamma_y^2+\gamma_z^2}.
\end{equation}
The Bloch phase boundaries are $M=M_{\pm}$, where the gap
closes at ${\boldsymbol k}=(0,0,0,0)$. When $a=0$, the spectrum for the lower band has doubly degeneracy and the model now still respects the CP combined symmetry, which acts on the Hamiltonian as
\begin{equation}
(CP)\mathcal{H}^T_{SI}(CP)^{-1}=-\mathcal{H}_{SI},~CP=\sigma_x\otimes\sigma_y,
\end{equation}
where $``T"$ denotes the  matrix transpose. The system hosts a $\mathbb{Z}_2$ topological invariant and thus $\mathcal{DD}=0$. When $a\neq 0$, the topology of this system belongs to $\mathbb{Z}$ class.  Note that the winding number of this NH system for all values of $a$ is $w=1$, which means it can not be used to characterize its topology.
%%%%%%%%%%%%%%%%%%%%%%%%%%%%%%%%%%%%%%%%%%%%%
\begin{figure}[htbp]\centering
\includegraphics[width=8.8cm]{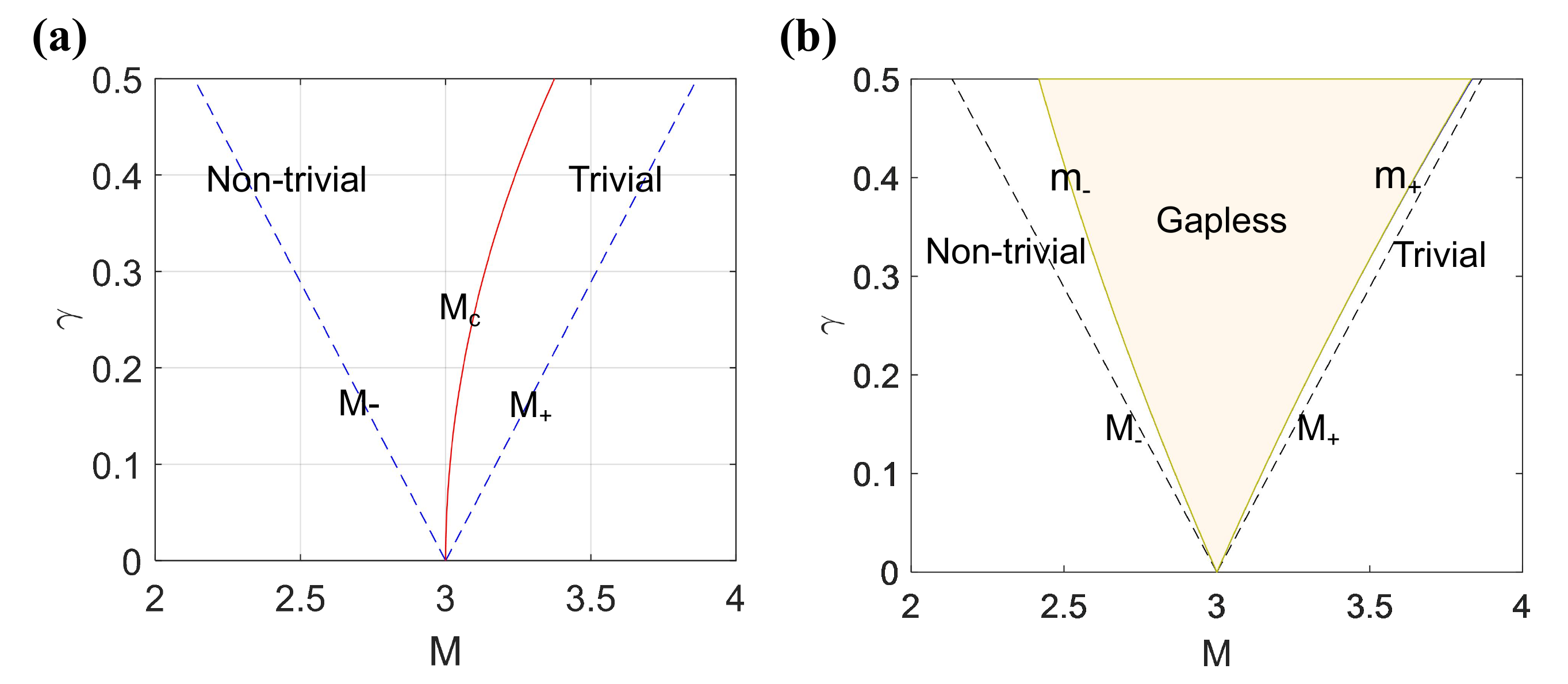}
 \caption{(Color online)  Phase diagram for OBC along (a) all the directions;(b) $z$ direction. (a)The phase boundary marked by red solid line is now modified by $M_c$. (b) The phase boundaries marked by red solid lines are modified by $m_{\pm}$. The dashed lines ($M_{\pm}$) denote the phase boundaries for the Bloch Hamiltonian under periodic boundary condition.} \label{Phasediagram}
\end{figure}
%%%%%%%%%%%%%%%%%%%%%%%%%%%%%%%%%%%%%%%%%%%%%%%
However, we can use the Abelian tensor Berry connection to identify its topological invariants as done in the 3D pseudo-Hermitian chiral topological insulator discussed in the main text.
We obtain that for $a\neq 0,\pm1$ ($a=\pm1$) the $\mathcal{H}_{SI}({\boldsymbol k})$-based $\mathcal{DD}$ invariant is $0$ for
$M>M_+$, $2$ ($1$) for $M<M_-$, and becomes non-definable in the
gapless region.
Differently from the previous 3D pseudo-Hermitian system, this model does not have the conventional bulk-edge correspondence which means the bulk Hamiltonian can not sufficiently
describe the topology of the system when we consider open boundary conditions in the real space.

%%%%%%%%%%%%%%%%%%%%%%%%%%%%%%%%%
\begin{figure}[htbp]\centering
\includegraphics[width=8.5cm]{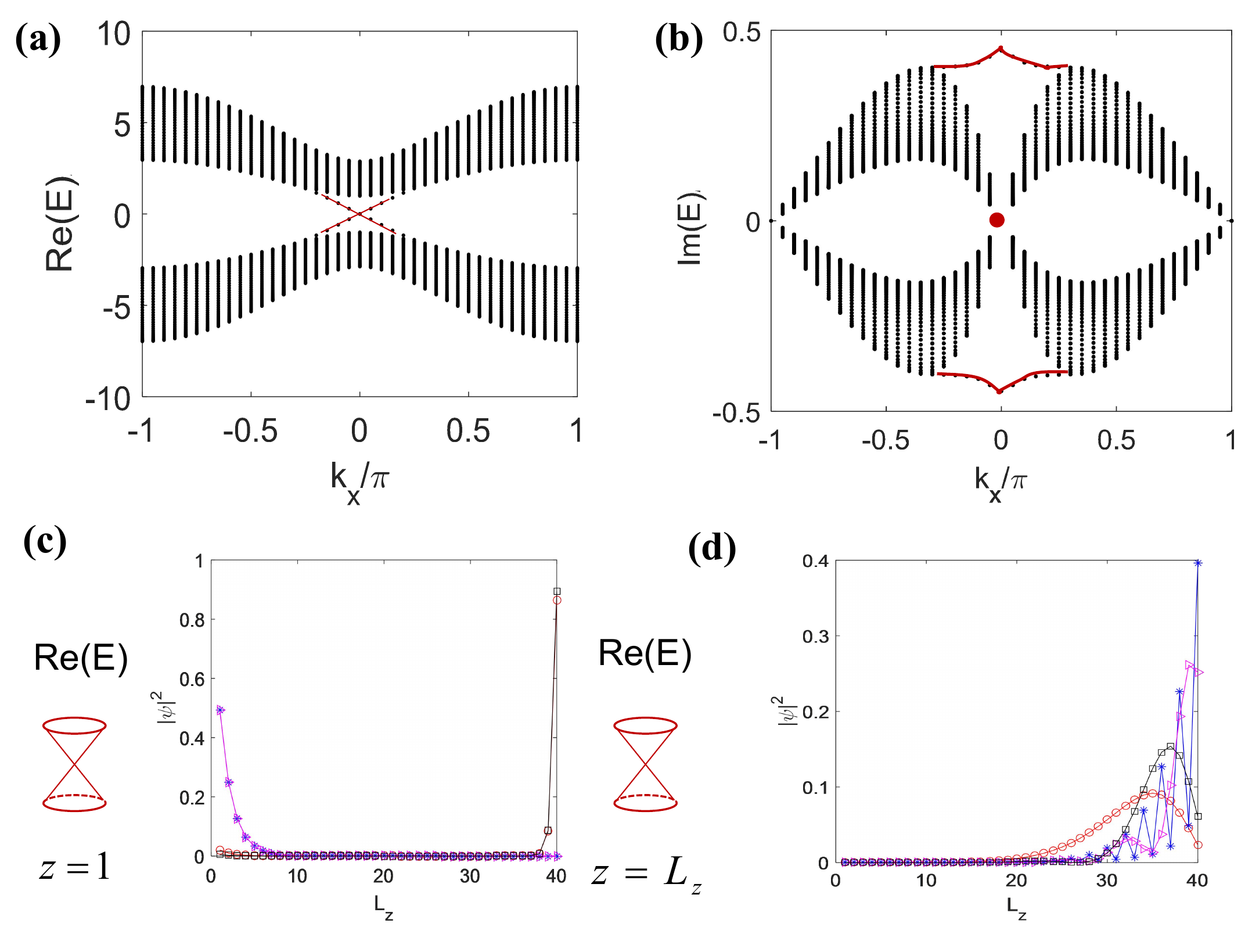}
 \caption{(Color online) Surface spectrum of $\mathcal{H}_{SI}$ with $k_y=0$ upon applying open boundary conditions along $z$ when $a=0$. (a)Real part of the surface spectrum;(b)Imaginary part of the surface spectrum. The surface modes ($k_x=0.1\pi$) are twofold degenerate colored in red located on each boundary, as shown in (c). (d) Skin modes (colored in black) in the bulk live in the right end. The parameters we used above are: $M=2.5$, $\gamma=0.2$, $J=1$, $L_z=40$.} \label{Z2phase}
 \end{figure}

\begin{figure}[htbp]\centering
\includegraphics[width=8.5cm]{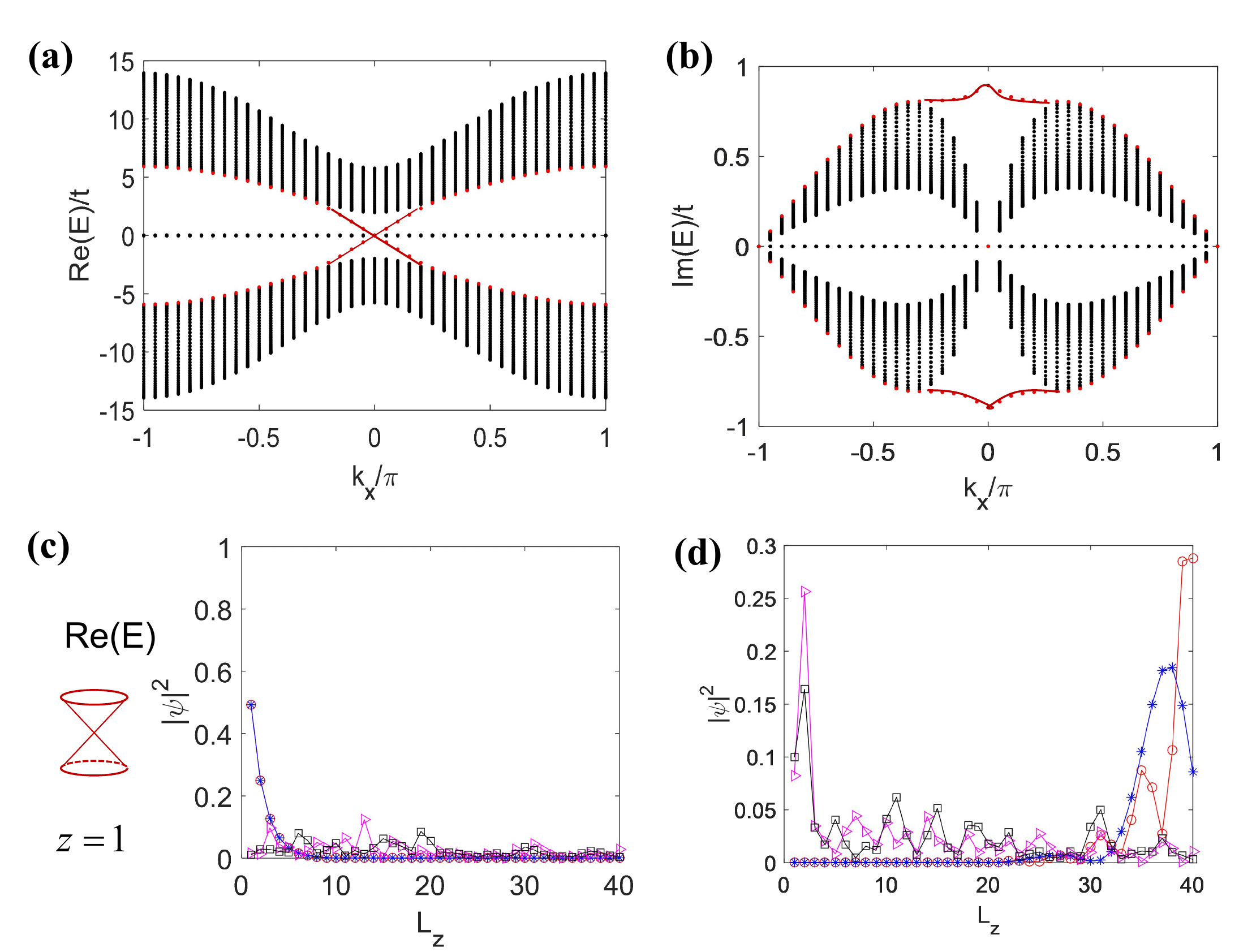}
 \caption{(Color online) Surface spectrum of $\mathcal{H}_{SI}$ with $k_y=0$ upon applying open boundary conditions along $z$ when $a=1$. (a)Real part of the surface spectrum;(b)Imaginary part of the surface spectrum. Only one surface mode ($k_x=0.1\pi$) survives in the left end, as shown in (c). (d) Skin modes which are part of the  bulk states live in the both left and right ends. The parameters we used above are: $M=2.5$, $\gamma=0.2$, $J=1$, $L_z=40$.} \label{Zphase2}
%\end{figure}
\end{figure}

 Similarly, we can use the non-Bloch theory based on complex-valued wave vectors to investigate its intriguing phase under OBC. The low-energy continuum model of Eq.~\eqref{SIHam} is given by
\begin{equation}
\begin{aligned}
\mathcal{H}(\boldsymbol k)=(k_x+i\gamma_x)\tilde{\Gamma}_x+(k_y+i\gamma_y)\tilde{\Gamma}_y+(k_z+i\gamma_z)\tilde{\Gamma}_z\\
+\left(M-3+\frac{k_x^2+k_y^2+k_z^2}{2}\right)\tilde{\Gamma}_w.
\end{aligned}
\end{equation}
To explore this non-Hermitian skin effect, we take a complex-valued momentum to describe open-boundary eigenstates: ${\boldsymbol k}\rightarrow \tilde{\boldsymbol k}+i\tilde{{\boldsymbol k}}'$, where the imaginary part $\tilde{\boldsymbol k}'$ takes the simple form $\tilde{k}_j'=-\gamma_j$ for small $\tilde{\boldsymbol k}$ in this model. Then we obtain the non-Bloch Hamiltonian as
\begin{equation}
\begin{aligned}
\tilde{\mathcal{H}}(\tilde{\boldsymbol k})&=\tilde{k}_x\tilde{\Gamma}_x+\tilde{k}_y\tilde{\Gamma}_y+\tilde{k}_z\tilde{\Gamma}_z+\\
&\left(\tilde{M}+\frac{\tilde{k}_x^2+\tilde{k}_y^2+\tilde{k}_z^2}{2}-i\sum_{j=x,y,z}\gamma_j\tilde{k}_j\right)\tilde{\Gamma}_w,
\end{aligned}
\end{equation}
where $\tilde{M}=M-3-(\gamma_x^2+\gamma_y^2+\gamma_z^2)/2$. When $\gamma_{x,y,z}=\gamma$,  the phase diagram with small $\gamma_j$ can be modified: topological nontrivial region $\tilde{M}<0$ becomes $M<3+\frac{3}{2}\gamma^2$, and the phase boundary: $M_c=3+\frac{3}{2}\gamma^2$, see Fig. \ref{Phasediagram}(a).
\begin{figure}
 \includegraphics[width=8.5cm]{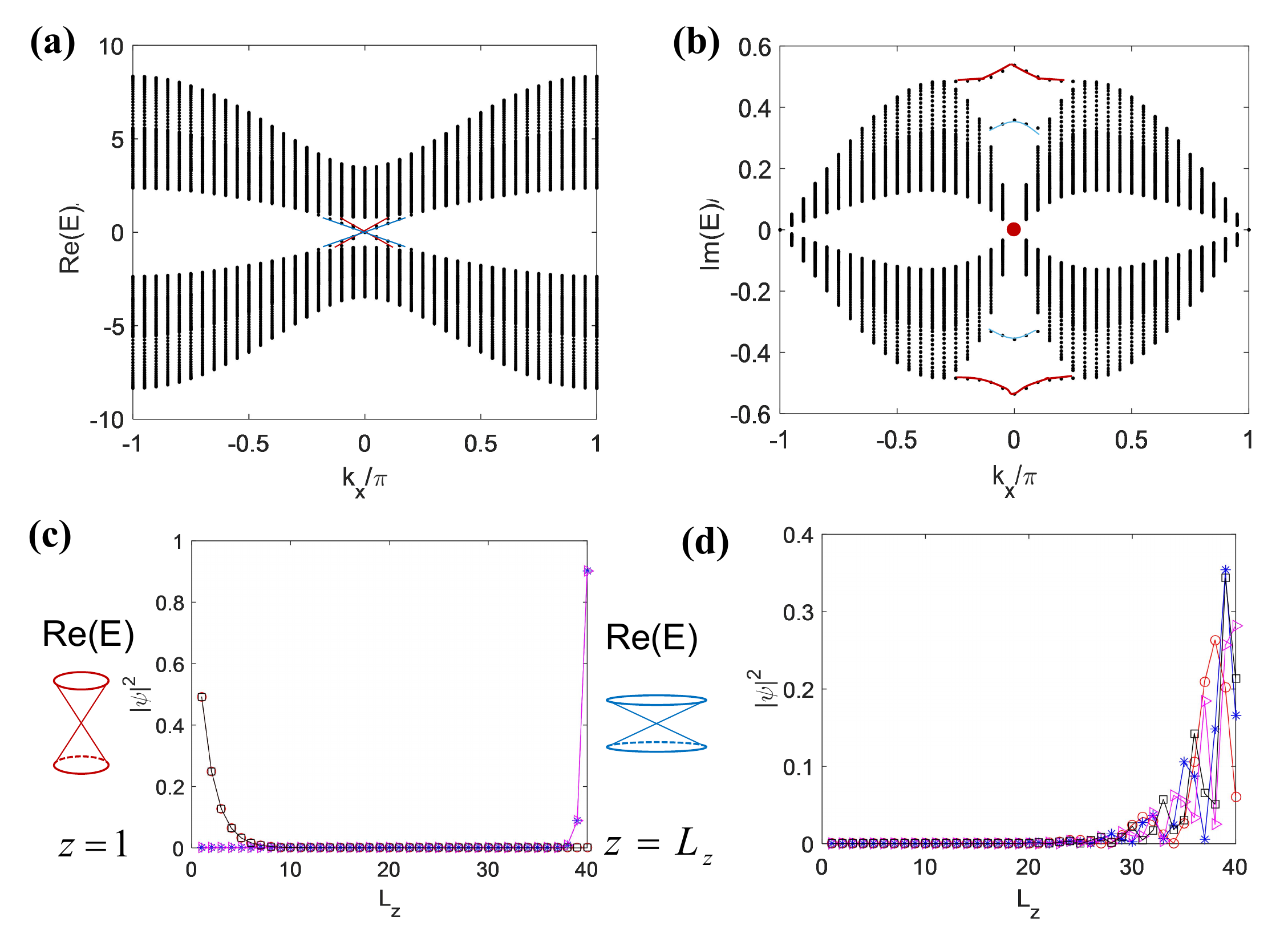}
 \caption{(Color online) Surface spectrum of $\mathcal{H}_{SI}$ with $k_y=0$ upon applying open boundary conditions along $z$ when $a=0.5$. (a)Real part of the surface spectrum;(b)Imaginary part of the surface spectrum. The degeneracy for the surface modes ($k_x=0.1\pi$) is then lifted upon increasing $a$, as shown in  (c). (d) Skin modes (colored in black) in the bulk live in the right end. The parameters we used above are: $M=2.5$, $\gamma=0.2$, $J=1$, $L_z=40$.} \label{Zphase1}
 \end{figure}
On the other hand, if we only consider the OBC along one direction (e.g., $z$ direction), then we have a different phase diagram. By replacing $k_z$ by $\tilde{k}_z-i\gamma$,  we obtain the corresponding non-Bloch Hamiltonian $\tilde{\mathcal{H}}_z(k_x,k_y,\tilde{k}_z)$. As shown in Fig.~\ref{Phasediagram}(b), the boundary phases are $m_{\pm}=3+\gamma^2/2\pm\sqrt{2}\gamma$ which are marked by red solid lines.  The topological nontrivial (trivial) region is given by $M<m_-$ ($M>m_+$) while the system is gapless in the region $M\in [m_-,m_+]$.
By considering the open boundary condition along $z$ direction with the lattice length $L_z$, we further explore the surface modes and the NH skin effects for this model. For $a=0$, the system is characterized by the winding number $w=1$ for the nontrivial region (here we take $M=2.5$).  Fig.\ref{Z2phase}(a-b) show that the surface spectra are twofold degenerate, as they describe the surface states on both boundaries (at $z=1$ and $z=L_z$), as sketched in Fig. \ref{Z2phase}(c). Fig. \ref{Z2phase}(d) shows there are skin modes in the bulk (black) live in the right end $(z=L_z)$.  As we increase $a$,  this degeneracy is then lifted, as shown in Fig. \ref{Zphase1}(a-c). The system with two occupied bands contributes $\mathcal{DD}=2$ and still supports the NH skin effects [Fig. \ref{Zphase1}(d)].  When reaching $a=1$ [Fig. \ref{Zphase2}(a-c)], the surface mode at $z=L_z$ vanishes into the zero-energy flat bulk band, while a single (non-degenerate) surface mode survives at $z=1$ characterized by $\mathcal{DD}=1$ for the lower occupied band. Now, only part of the bulk states are the skin modes which live in the both sides[Fig. \ref{Zphase2}(d)].

 %%%%%%%%%%%%%%%%%%%%%%%%%%%%%%%%%%%%%%%%%%%%%%%
%%%%%%%%%%%%%%%%%%%%%%%%%%%%%%%%%%%%%%%%%%%%%%%

\section{Dilated Hamiltonian in experiment}\label{Experiment}
\begin{figure*}[htbp]
\centering
    \includegraphics[width=14cm]{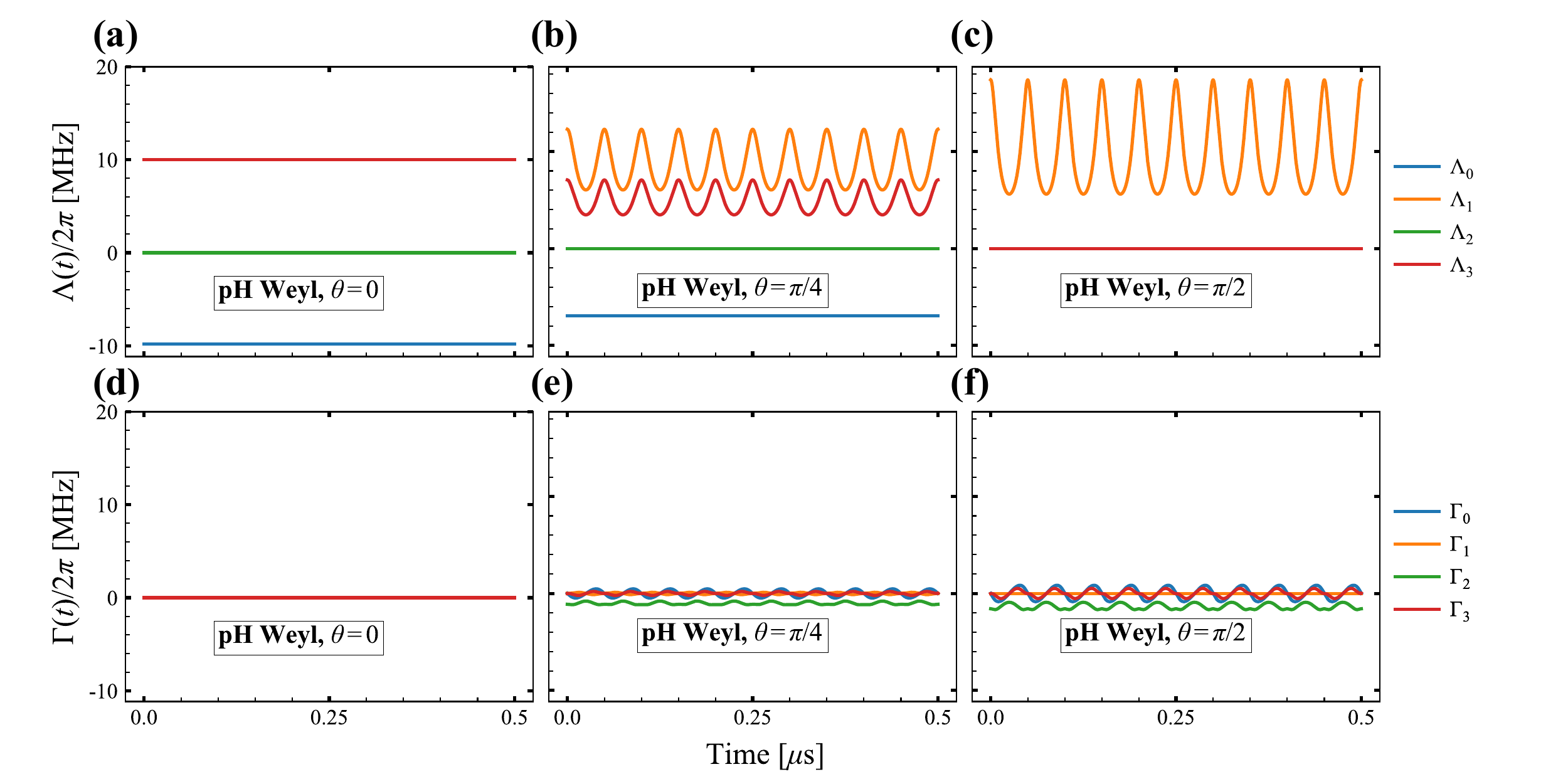}
    \caption{
        \textbf{Parameters varying with time.}
        The parameters designed in the dilated Hamiltonian $H_{s,a}(t)$ for a pH Weyl monopole as a function of evolution time $t$
        with $k/2\pi = 10$ [MHz], $q = 10, \varphi=0$ for
        \textbf{(a, d)} $\theta = 0$,
        \textbf{(b, e)} $\theta = 0.25 \pi$, and
        \textbf{(c, f)} $\theta = 0.5 \pi$.
        \label{fig:FIG1}
        }\label{parameterevolution_pHWeyl}
\end{figure*}

Let us now consider a 3D pH Weyl Hamiltonian (\ref{pHWeyl}) in  the (ellipsoidal) parameter space
\begin{equation}
    H_s=\tilde{\mathcal{H}}_W = \frac{d k}{c}\cos{\theta} \sigma_0 + k
        \left(
          \begin{array}{cc}
          \cos{\theta} & \sqrt{\frac{a}{b}}\sin{\theta} e^{-i \varphi}\\
          \sqrt{\frac{b}{a}}\sin{\theta} e^{i \varphi} & - \cos{\theta}\\
          \end{array}
        \right),
\end{equation}
with $k = \sqrt{c^2 k_z^2 + ab(k_x^2 + k_y^2)}$.
By employing the Naimark-Dilation approach, we can extend $H_s$ into a dilated Hermitian system $H_{s,a}$ with two ancilla qubits \cite{Gunther2007,Gunther2008}.
The corresponding eigenvector is written as
\begin{equation}
\begin{aligned}
    |\Psi(t) \rangle &= \alpha(t)\left(
        \begin{matrix}
            |\psi(t)\rangle\\
            |\chi(t)\rangle
        \end{matrix}\right)\\ &=\alpha(t)\left(|0\rangle_a\otimes|\psi(t)\rangle+|1\rangle_a\otimes|\chi(t)\rangle\right),
\end{aligned}
\end{equation}
where $\alpha(t) = \frac{1}{\sqrt{\langle \psi(t)|(1+\eta^{-2})|\psi(t)\rangle}}$ is the normalization coefficient, $|\chi(t)\rangle=\eta^{-1}|\psi(t)\rangle$ is related to the pH evolution state $|\psi(t)\rangle$ with
\begin{equation}
    \eta^{-1} = \left(
                  \begin{array}{cc}
                    \sqrt{\frac{b}{a}} & 0 \\
                    0 & \sqrt{\frac{a}{b}} \\
                  \end{array}
                \right)
\end{equation}
which is a Hermitian operator and satisfies the pH condition $H_s\eta = \eta H_s^{\dagger}$.
Therefore, the dilated Hamiltonian \cite{Gunther2008,YChu2020} is given by
\begin{equation}\label{DH1}
    H_{s,a} = \sigma_0^a \otimes \Lambda + \sigma_y^a \otimes \Gamma
\end{equation}
with
\begin{equation}
    \begin{aligned}
        \Lambda&=\beta^{-1} (H_s\eta + \eta^{-1} H_s) \sigma_0\\
        &=\frac{dk}{c}\cos{\theta} +k \left[\cos{\theta} \sigma_z + \frac{\sqrt{2ab}}{a + b}\sin{\theta} (\sigma_x \cos\varphi+ \sigma_y \sin\varphi)\right]\\
         \Gamma&=i\beta^{-1}(H_s - H_s^{\dagger})= k\frac{a - b}{a + b} \sin{\theta} (\sigma_x \sin \varphi- \sigma_y \cos\varphi).
    \end{aligned}
\end{equation}
Meanwhile, our pH Hamiltonian can be demonstrated on the Hermitian system with the extended Naimark-Dilated method \cite{YWu2019}.
For instance, by introducing a dilated state
\begin{equation}
|\Psi(t)\rangle = \alpha(t) \left(|+\rangle\otimes|\psi(t)\rangle + \eta^{-1}(t) |-\rangle\otimes|\psi(t)\rangle\right)
\end{equation}
and choosing an appropriate parameter $\eta(t=0) = \eta_0 \sigma_0$ to satisfy that the dilating Hamiltonian keeps Hermiticity under evolution,
and calculating
\begin{equation}
M(t) = \eta^{-2}(t) + \sigma_0 = \mathcal{T} e^{-i\int_0^t H_s^{\dagger}(t')dt'} M(0) \overline{\mathcal{T}} e^{i\int_0^t H_s(t')dt'},
\end{equation}
 we have
\begin{equation}\label{DH2}
    H_{s,a}(t)=\sigma_0^a\otimes \Lambda(t) + \sigma_3^a \otimes \Gamma(t)
\end{equation}
with
\begin{equation}
    \begin{array}{cc}
        \Lambda(t) =  \{ H_s(t) + [i\frac{d\eta^{-1}(t)}{dt} + \eta^{-1}(t) H_s(t)]\eta^{-1}(t)\}M^{-1}(t)\\
        \Gamma(t)  =  i[H_s(t) \eta^{-1}(t) - \eta^{-1}(t)H_s(t) - i\frac{d\eta^{-1}(t)}{dt}]M^{-1}(t)\\
    \end{array},
\end{equation}
where $|-\rangle = \frac{1}{\sqrt{2}}\left(|0\rangle_{a} - i|1\rangle_{a}\right)$ and $|+\rangle = \frac{-i}{\sqrt{2}}\left(|0\rangle_{a} + i|1\rangle_{a}\right)$ are the eigenstates of $\sigma_y$ of the ancilla qubit.
By decomposing
$\Lambda_i(t) = {\rm Tr}[\Lambda(t) \sigma_i]/2$,
$\Gamma_i(t)  = {\rm Tr}[\Gamma (t) \sigma_i]/2$ in terms of Pauli operators
$\sigma_{i = 0, 1, 2, 3} = \sigma_0, \sigma_x, \sigma_y, \sigma_z$,
it shows that constructing the dilated Hermitian Hamiltonian $H_{s,a}(t)$ of an arbitrary 3D pH Weyl Hamiltonian needs to design eight real parameters, as shown in Fig.~\ref{parameterevolution_pHWeyl}.
%%%%%%%%%%%%%%%%%%%%%%%%%%%%%%%%%%%%%%%%%%%%%%%

\begin{figure*}[htbp]
\centering
    \includegraphics[width=14cm]{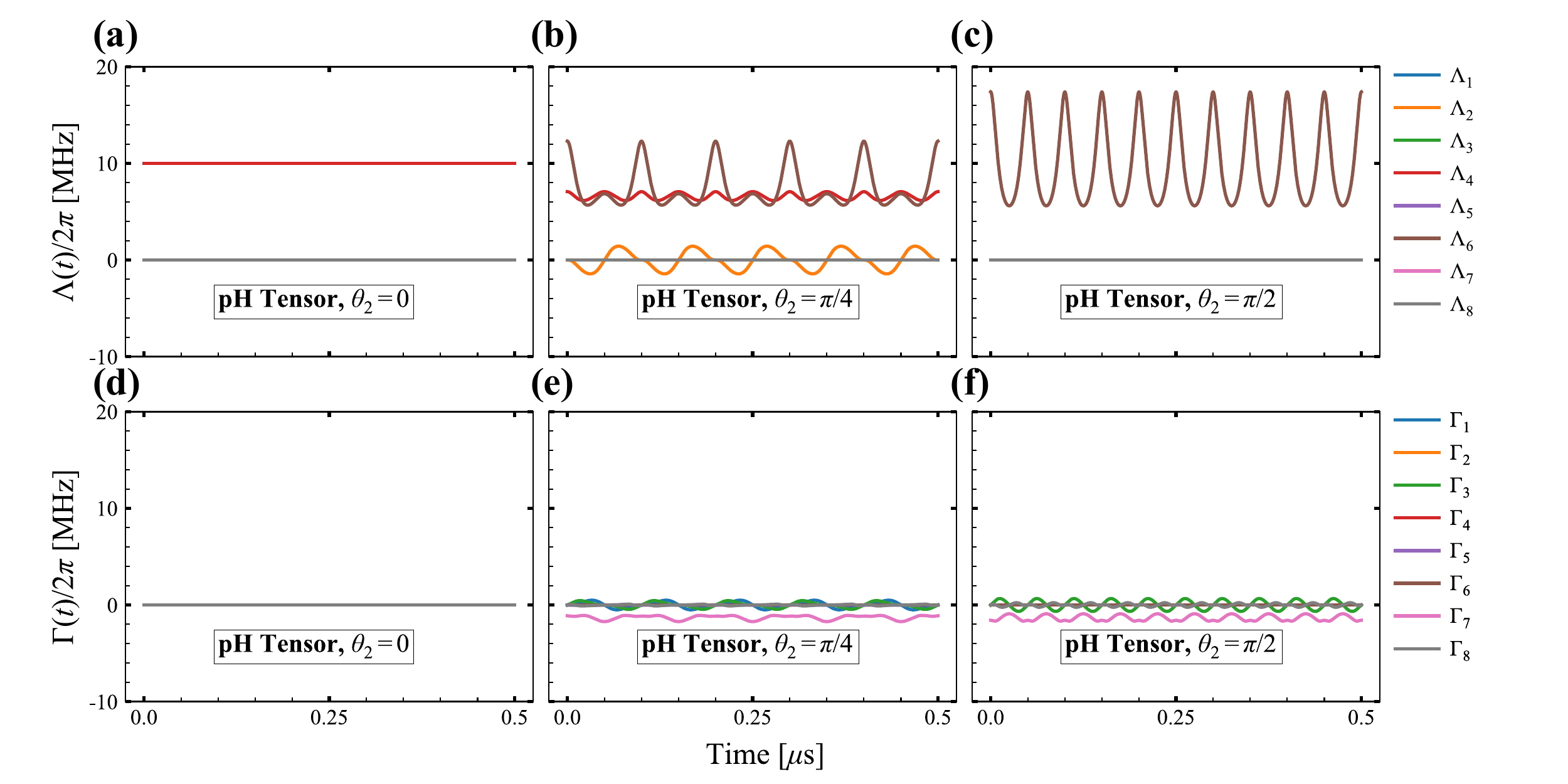}
    \caption{
        \textbf{Parameters varying with time.}
        The parameters designed in the dilated Hamiltonian $H_{s,a}(t)$ for a pH tensor monopole as a function of evolution time $t$
        with $k/2\pi = 10$ [MHz], $q = 10, \theta_1=0.5\pi, \varphi=0$ for
        \textbf{(a, e)} $\theta_2 = 0$,
        \textbf{(b, d)} $\theta_2 = 0.25 \pi$, and
        \textbf{(c, f)} $\theta_2 = 0.5 \pi$, which are correspond to the pH tensor Hamiltonian
        reduce to only Hermitian evolution $|0\rangle_{s} \rightleftharpoons |2\rangle_{s}$, mixed evolution $|0\rangle_{s} \rightleftharpoons |2\rangle_{s} \rightleftharpoons |1\rangle_{s}$ and only pH evolution $|1\rangle_{s} \rightleftharpoons |2\rangle_{s}$, respectively. Note that $|0\rangle_{s}=(1,0,0)^T$,$|1\rangle_{s}=(0,1,0)^T$,and $|2\rangle_s=(0,0,1)^T$.
        \label{fig:FIG1}
        }\label{parameterevolution}
\end{figure*}
%%%%%%%%%%%%%%%%%%%%%%%%%%%%%%%%%%%%%%%%
\begin{figure*}[htbp]
\centering
    \includegraphics[width=14cm]{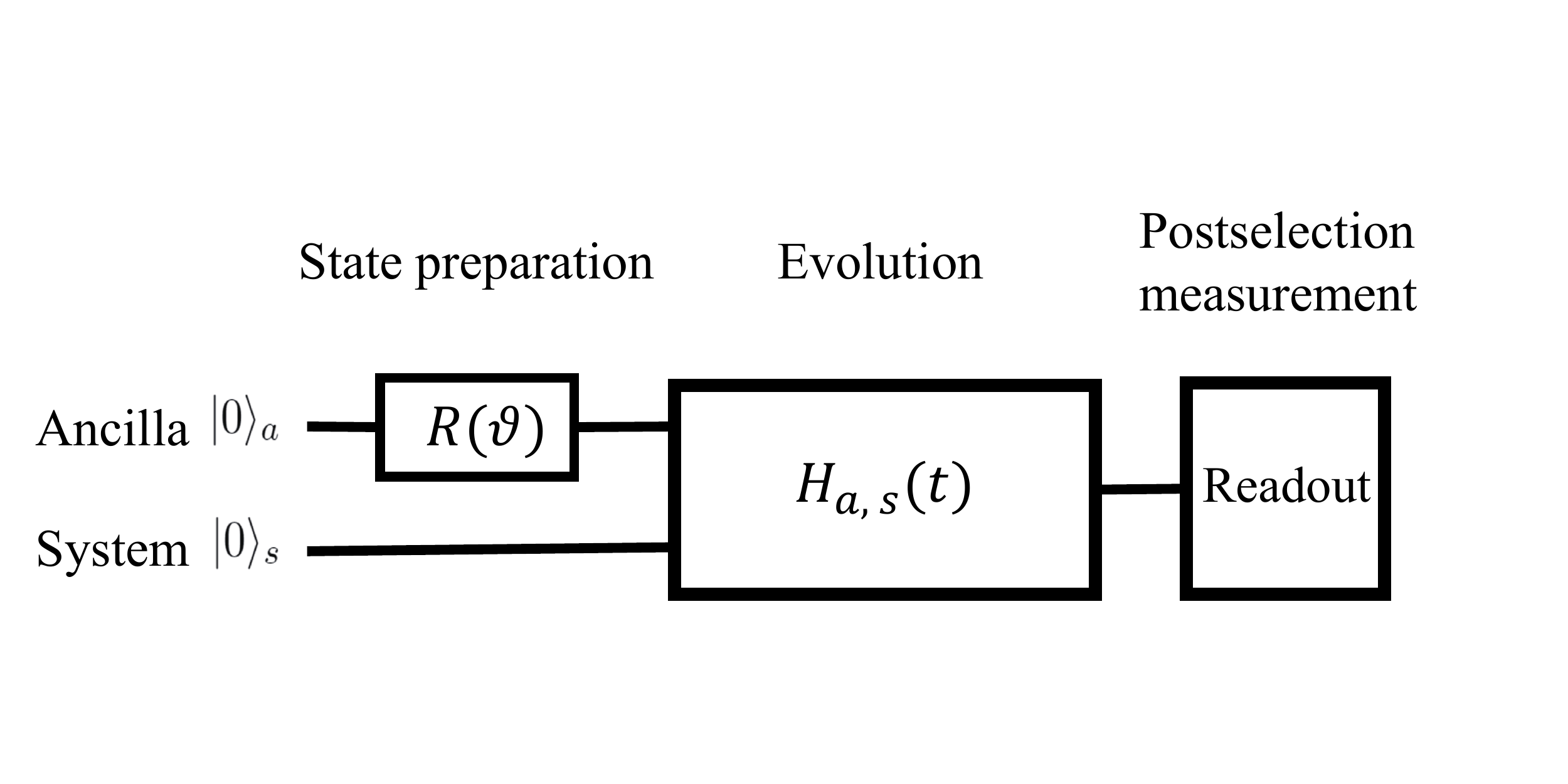}
    \caption{
        \textbf{The experiment schematic diagram of (extendted) Naimark-Dilated protocol.}
        $R(\vartheta)$ denotes the rotation in ancilla system for preparing the state $|\Psi(t=0)\rangle$.
        Then by modulating experiment parameters (such as $q$, $k$, $\theta$, $\varphi$), the Hermitian system
        evolves under dilated Hamiltonian $H_{s,a}(t)$. Finally, the effective pH system
        dynamic evolution is acquired by the postselection measurement.
    }\label{Schematicdiagram}
    \includegraphics[width=14cm]{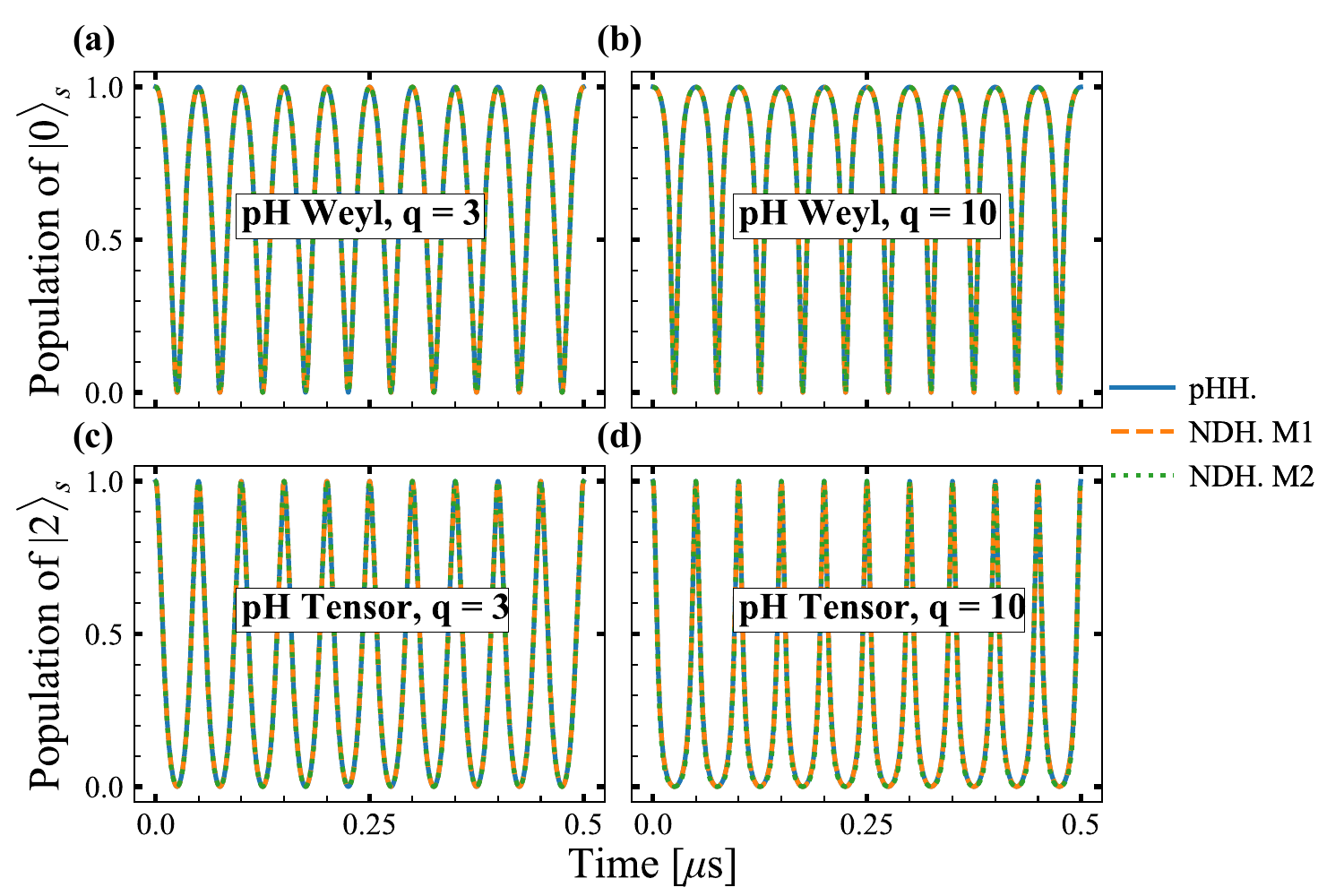}
    \caption{
        \textbf{Evolution simulation.}
        The blue line is calculated from the pseudo-Hermitian Hamiltonian (pHH),
        the orange dashed line and the green dotted line are calculated from the Naimark-Dilated Hamiltonian [Eq. \eqref{DH1}] (NDH. M1) and the extended Naimark-Dilated Hamiltonian [Eq. \eqref{DH2}] (NDH. M2), respectively.
        For \textbf{(a)}, the dynamic evolution of the 3D pH Weyl Hamiltonian, we set $k/2\pi =10$ [MHz], $q=3$, $\theta=0.5 \pi$, $\phi=0$,
        while for \textbf{(b)}, they are the same as for \textbf{(a)} except $q=10$.
        For the dilated pH tensor Hamiltonian simulation, we set $k/2\pi =10$ [MHz], $\theta_1=0$, $\theta_2 = 0.5 \pi$, $\phi=0$,
        with \textbf{(c)} $q=3$, while for \textbf{(d)} $q=10$. Here, the initial state is prepared on $|2\rangle_s$ for the convenience of studying the evolution in different parameters.
        }\label{Stateevolution}
\end{figure*}

In a similar way, we can consider a 4D pH tensor monopole described in Eq.~\eqref{pHTM} in 4D (hyper-ellipsoidal) parameter space,
\begin{equation}
\begin{aligned}
    H_s=\tilde{\mathcal{H}}_{T}=
 k_x\lambda_4 + \lambda_5 k_y + \tilde{\lambda}_6 k_z+\tilde{\lambda}_7 k_w= \\ \nonumber
\left(
\begin{array}{ccc}
 0 & 0 & k_x-ik_y \\
 0 & 0 & a(k_z-ik_w) \\
 k_x+ik_y & b(k_z+ik_w) & 0 \\
 \end{array}
 \right),
    \end{aligned}
\end{equation}
with $k = \sqrt{k_x^2 + k_y^2 + ab(k_z^2 + k_w^2)}$. After performing the parameterization as in Eq. \eqref{PTransform}, and  doing some straightforward calculations, we obtain a similar extended Hamiltonian that takes the form in \eqref{DH1} and \eqref{DH2}, respectively. The corresponding decomposed matrices are given by
\begin{equation}
    \begin{aligned}
        \Lambda &= k\cos{\theta_1}\lambda_4 + k\sin{\theta_1}\cos\theta_2\lambda_5\\
         &+ k\frac{2\sqrt{ab}}{a+b} \sin\theta_1\sin\theta_2(\cos{\varphi}\lambda_6 + \sin{\varphi}\lambda_7),\\
       \Gamma &=k \frac{a-b}{a+b}\sin\theta_1\sin\theta_2(\sin{\varphi}\lambda_6 - \cos{\varphi}\lambda_7),\\
    \end{aligned}
\end{equation}
and
\begin{equation}\label{DH3}
    \begin{array}{cc}
        \Lambda_i(t) = {\rm Tr}[\Lambda(t) \lambda_i]/2,~~
        \Gamma_i(t)  = {\rm Tr}[\Gamma (t) \lambda_i]/2
    \end{array}
\end{equation}
in terms of Gell-Mann matrices $\lambda_{i \in [1, 8]}$.
As for Eq.~\eqref{DH3}, the numerical calculations show that there are sixteen real parameters %$\left[\Lambda_{i=2,4,6}(t), \Gamma_{i=1,3,5,7,8}(t)\right]$
need to be designed in order to obtain the last dilated Hamiltonian in Eq.~\eqref{DH2}. For instance, they are varying with evolution time $t$ as shown in Fig.~\ref{parameterevolution} under some experimental conditions.
Since the corresponding dilated Hermitian systems can be conveniently implemented using superconducting quantum circuits, nitrogen-vacancy center in diamond and trapped ions, etc.,
the proof-of-principle verification of our protocol is feasible by post-selection measurement which is a normal technique in quantum measurement.
Here we present the experiment circuit diagram, as depicted in Fig.~\ref{Schematicdiagram}.

To verify that our pH Hamiltonian can be emulated on the dilated Hermitian systems [Eq. \eqref{DH1} and Eq. \eqref{DH2}],
we carry out numerical simulations of dynamic evolutions, respectively.
Our results confirm that the two Hermitian systems respectively constructed by employing the two methods can well describe our proposed pH Hamiltonian, as shown in Fig.~\ref{Stateevolution}.

\section{Detecting the NH quantum geometry tensor through Rabi oscillation}\label{Dection}
\begin{figure*}[htbp]\centering
\includegraphics[width=15cm]{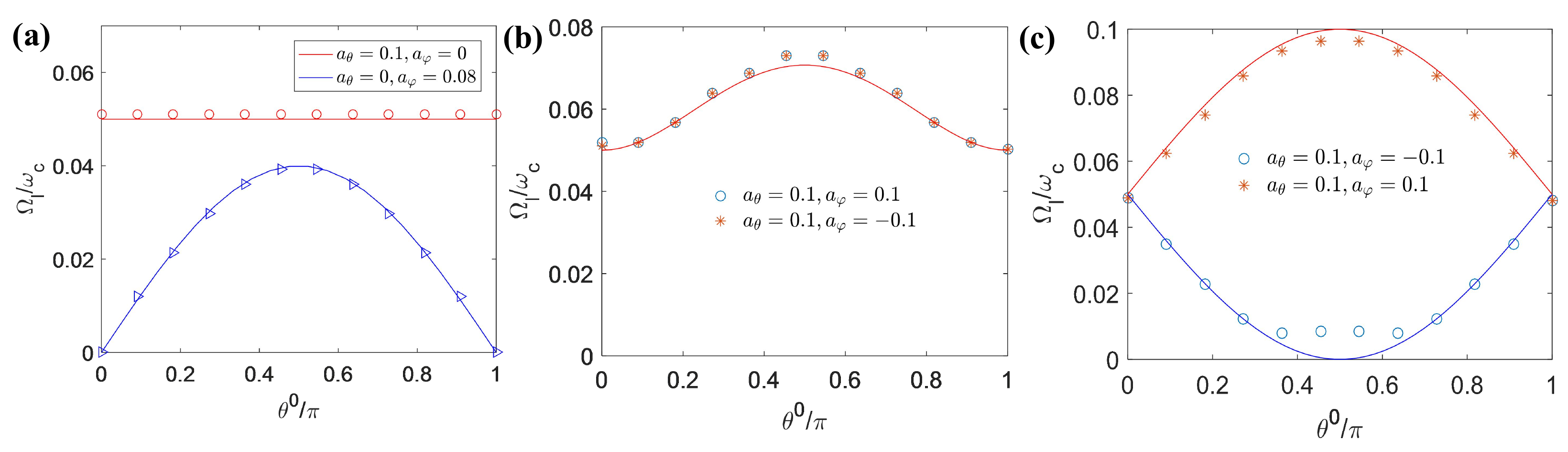}
\includegraphics[width=11cm]{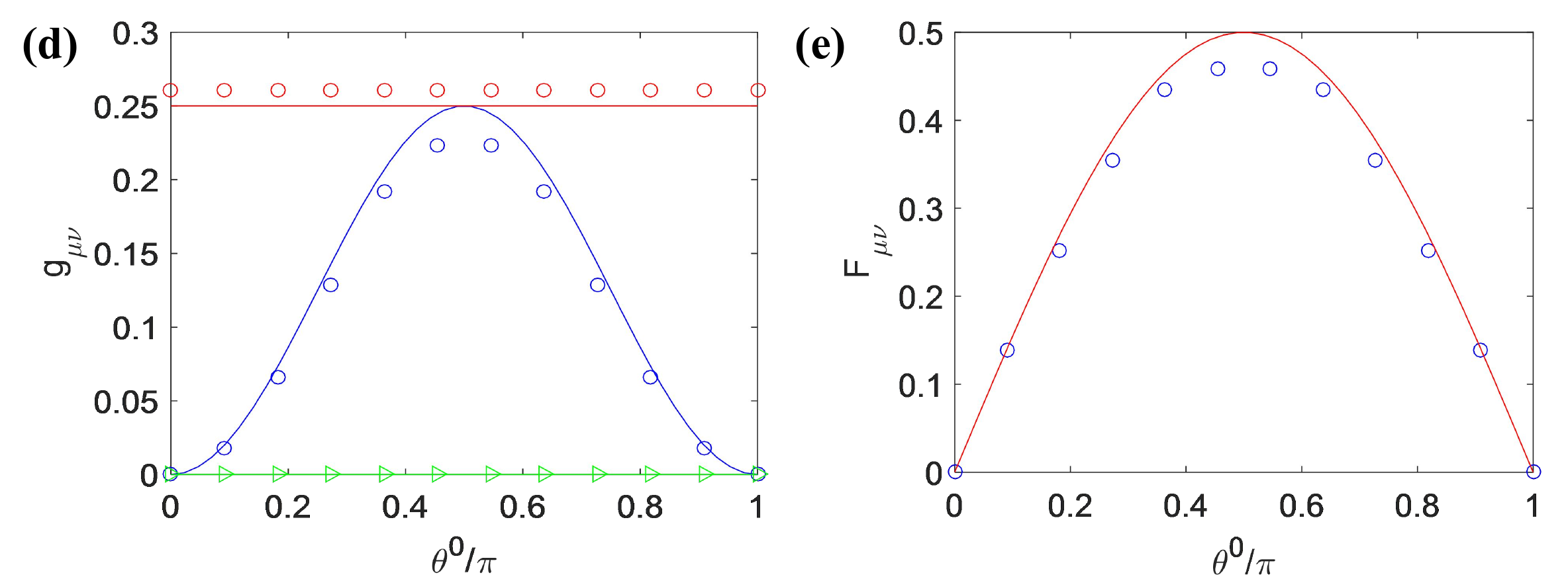}
 \caption{(Color online) Coherent transitions induced by parametric modulations. Resonant oscillation under the parametric modulations: (a-b)$a_{\theta}=0.1$ ($a_{\theta}=0$), $a_{\varphi}=0$ ($a_{\varphi}=0.08$); (c) $a_{\theta}=0.1$, $a_{\varphi}=0.1$. The other parameters we used are: $a=3$,$b=1$, $E_0=20.98$ MHz. The solid curves show theoretical predictions presented in Eq. \eqref{QMT}. In (a-e), we set $\varphi^0=0$.} \label{RF}
\end{figure*}

In this section, we provide the measurement protocol to extract the NH (Abelian) quantum metric and Berry curvature in pH systems.
For $N$-level systems in the presence of pseudo-Hermiticity, we find the following relations,
\begin{equation}\label{pHrelation}
\begin{aligned}
\eta H^{\dagger} \eta^{-1}&=H,~H^{\dagger}=\eta^{-1}H\eta,~\eta^{\dagger}=\eta,~(\eta^{-1})^{\dagger}=\eta^{-1},\\
H|u_n^R\rangle&=E_n|u_n^R\rangle,~H^{\dagger}|u_n^L\rangle=E_n|u_n^L\rangle,~E_n=E_n^*,\\
L^{\dagger}R=R^{\dagger}L&=1,R=(|u_1^R\rangle,...,|u_N^R\rangle), L=(|u_1^L\rangle,...,|u_N^L\rangle),\\
&|u_n^R\rangle=\eta|u_n^L\rangle,~|u_n^L\rangle=\eta^{-1}|u_n^R\rangle.
\end{aligned}
\end{equation}
Based on the above relations, the NH QGT can now be rewritten as
\begin{equation}
\begin{aligned}
Q_{\mu\nu}^n&=\frac{1}{2}\left(\langle\partial_{\mu}u_n^L|(1-P_n)|\partial_{\nu}u_n^R\rangle+\langle\partial_{\mu}u_n^R|(1-P_n^{\dagger})|\partial_{\nu}u_n^L\rangle\right)\\
&=\langle\partial_{\mu}u_n^L|(1-P_n)|\partial_{\nu}u_n^R\rangle
=\sum_{n\neq m}\langle\partial_{\mu}u_n^L|u_m^R\rangle\langle u_m^L|\partial_{\nu}u_n^R\rangle\\
&=\sum_{n\neq m}\frac{\langle u_n^L|\partial_{\mu}H|u_m^R\rangle\langle u_m^L|\partial_{\nu}H|u_n^R\rangle}{(E_n-E_m)^2}\\
&=\sum_{n\neq m}\frac{\langle u_n^L|\partial_{\mu}H|u_m^R\rangle\langle u_m^R|\partial_{\nu}H^{\dagger}|u_n^L\rangle}{(E_n-E_m)^2},
\end{aligned}
\end{equation}
where we have used the relation $\langle u_n^L|\partial_{\mu}u_m^R\rangle (E_n-E_m)=\langle u_n^L|\partial_{\mu}H|u_m^R\rangle$ $(n\neq m)$ in the last step.
Inspired by the idea on extracting the QGT through the Rabi oscillations \cite{MYu2020}, we here extend this method to pH systems. For simplicity, we first consider a generic pH two-level Hamiltonian in the ellipsoidal coordinates expressed as
\begin{equation}\label{pHHam}
\begin{aligned}
H(\theta,\varphi)&=\frac{E_0}{2}\left(
\begin{array}{cc}
 \cos\theta& \sqrt{{\frac{a}{b}}}\sin\theta e^{-i\varphi} \\
\sqrt{{\frac{b}{a}}}\sin\theta e^{i\varphi} & -\cos\theta \\
\end{array}
\right)\\
&=E_1|u_1^R\rangle\langle u_1^L|+E_2|u_2^R\rangle\langle u_2^L|,
\end{aligned}
\end{equation}
where $E_{1,2}=\pm E_0$ denotes the eigenvalue, the model respects pseudo-Hermiticity: $\eta H^{\dagger}\eta^{-1}=H$ with $\eta=\text{diag}(a,b)/\sqrt{ab}$.
Below we consider two types of parametric modulation $[\lambda_{\mu}(t),\lambda_{\nu}(t)]\leftrightarrow[\theta(t),\varphi(t)]$:
(a) the linear parametric modulation
with $\lambda_{\mu}(t)=\lambda_{\mu}^0+a_{\mu}\sin\omega t$, $\lambda_{\nu}(t)=\lambda_{\nu}^0+a_{\nu}\sin\omega t$;
(b) the elliptical parametric modulation with  $\lambda_{\mu}(t)=\lambda_{\mu}^0+a_{\mu}\sin\omega t$, $\lambda_{\nu}(t)=\lambda_{\nu}^0+a_{\nu}\cos\omega t$.
To extract the diagonal component of the quantum metric, we consider a weakly linear parametric modulation with $a_{\mu}\neq 0$ and $a_{\nu}=0$ when $|a_{\mu}|\ll 1$, the time-dependent Hamiltonian is expanded as
\begin{equation}
H[{\boldsymbol \lambda}(t)]=H({\boldsymbol \lambda}^0)+a_{\mu}\sin \omega t\partial_{\mu}H({\boldsymbol \lambda}^0).
\end{equation}
After performing a unitary transformation $L^{\dagger}H[{{\boldsymbol \lambda}(t)}]R$ and then the rotating wave approximation, we obtain
\begin{equation}
H_{rot}[{\boldsymbol \lambda}^0]=
\left(
  \begin{array}{cc}
    E_1-\omega & \Omega_{12} \\
    \Omega^*_{12} & E_2 \\
  \end{array}
\right),
\end{equation}
where $|E_1-E_2|=\omega_c$, and
\begin{equation}
\Omega_{12}=\frac{a_{\mu}}{2}\langle u_1^L|\partial_{\mu}H({\boldsymbol \lambda}^0)|u_2^R\rangle.
\end{equation}
Thus, we obtain the following relation
\begin{equation}
g^n_{\mu\mu}=\sum_{n\neq m}\frac{|\langle u_n^L|\partial_{\mu}H|u_m^R\rangle|^2}{(E_n-E_m)^2}=\sum_{n\neq m}\frac{4|\Omega_{nm}|^2}{a^2_{\mu}\omega_{nm}^2}=\frac{4|\Omega_{12}|^2}{a^2_{\mu}\omega_c^2}.
\end{equation}
For the resonant case when $\omega=\omega_c$, the Rabi frequency of this coherence transition is
\begin{equation}
\Omega_l=2|\Omega_{12}|=a_{\mu}\sqrt{g_{\mu\mu}^n}\omega_c.
\end{equation}
It is worth to notice that the Rabi oscillation between these two bi-orthogonal states determined by
\begin{equation}\label{Probability}
|c_n(t)|^2=|\langle u^L_n(0)|\psi^R(t)\rangle|^2=|\langle u^R_n(0)|\psi^L(t)\rangle|^2.
\end{equation}
In principle, to obtain the bi-orthogonal eigenvectors, $|\psi^{L(R)}\rangle$, we need to do the implementation on the Hamiltonian $H$ and  $H^{\dagger}$ simultaneously for the generic NH system. Fortunately, due to the present of pseudo-Hermiticity in our model, we just need to do the experiment in the system $H[{\boldsymbol \lambda}(t)]$. By employing the relations mentioned in Eq.~\eqref{pHrelation}, we can easily prove that
\begin{equation}
\begin{aligned}
|u^L_n(t)\rangle&=\mathcal{T}e^{-i\int_0^t H^{\dagger}[{\boldsymbol\lambda}(t')]dt'}|u^L_n(0)\rangle\\
&=\mathcal{T}\eta^{-1}\eta e^{-i\int_0^t H^{\dagger}[{\boldsymbol\lambda}(t')]dt'}\eta^{-1}\eta|u^L_n(0)\rangle\\
&=\eta^{-1}\mathcal{T}e^{-i\int_0^t H[{\boldsymbol\lambda}(t')]dt'}|u_n^R(0)\rangle=\eta^{-1}|u_{n}^R(t)\rangle.
\end{aligned}
\end{equation}
Experimentally, we can start with an initial state $|\psi^{R}(0)\rangle=|u_{1}^{R}(0)\rangle$ prepared at the lower band of $H({\boldsymbol \lambda}^0)$. By substituting the above relation into Eq.~\eqref{Probability}, we get the observable quantity expressed as
\begin{equation}
|c_n|^2=|\langle u^R_n(0)|\eta^{-1}|u_1^R(t)\rangle|^2,
\end{equation}
where we can extract the Rabi frequency $\Omega_l$.
Next, we  proceed to general case for the linear modulation with $a_{\mu}\neq 0$ and $a_{\nu}\neq 0$, meanwhile, $|a_{\mu(\nu)}|\ll 1$. The time-dependent Hamiltonian is expanded as
 \begin{equation}
 H[{\boldsymbol \lambda}(t)]= H({\boldsymbol \lambda}^0)+a_{\mu}\sin \omega t\partial_{\mu}H({\boldsymbol \lambda}^0)+a_{\nu}\sin \omega t\partial_{\nu}H({\boldsymbol \lambda}^0).
 \end{equation}
The coherent transition Rabi frequency is
\begin{equation}
\Omega_{12}=\frac{1}{2}\langle u_1^L|a_{\mu}\partial_{\mu}H({\boldsymbol \lambda}^0)+a_{\nu}\partial_{\nu}H({\boldsymbol \lambda}^0)|u_2^R\rangle,
\end{equation}
and the related resonant Rabi frequency for this two-level model is $\Omega_l=2|\Omega_{12}|$. In a similar way, we obtain the relation like in the Hermitian case \cite{MYu2020}, expressed as
\begin{equation}
\sum_{n\neq m}\frac{4|\Omega_{nm}|^2}{\omega_{nm}^2}=\frac{\Omega_{l}^2}{\omega_c^2}=a^2_{\mu}g^n_{\mu\mu}+2a_{\mu}a_{\nu}g^n_{\mu\nu}+a^2_{\nu}g^n_{\nu\nu},
\end{equation}
which leads to the following expression
\begin{equation}
g^n_{\mu\nu}=[\Omega_l(a_{\mu,\nu})^2-\Omega_l(a_{\mu},-a_{\nu})^2]/\left(4a_{\mu}a_{\nu}\omega_c^2\right).
\end{equation}
So far, we have extracted the quantum metric tensor through the linear modulation. To extract the imaginary part of the NH QGT, i.e., the Berry curvature, we should consider the elliptical parametric modulation with  $\lambda_{\mu}(t)=\lambda_{\mu}^0+a_{\mu}\sin\omega t$, $\lambda_{\nu}(t)=\lambda_{\nu}^0+a_{\nu}\cos\omega t$.  After implementing a similar derivation, we have the relation connecting the Berry curvature and the corresponding Rabi frequency,
\begin{equation}
\sum_{n\neq m}\frac{4|\Omega_{nm}|^2}{\omega_{nm}^2}=\frac{\Omega_{c}^2}{\omega_c^2}=a^2_{\mu}g^n_{\mu\mu}+a_{\mu}a_{\nu}F^n_{\mu\nu}+a^2_{\nu}g^n_{\nu\nu},
\end{equation}
which leads to
\begin{equation}
F^n_{\mu\nu}=[\Omega_c(a_{\mu,\nu})^2-\Omega_c(a_{\mu},-a_{\nu})^2]/\left(2a_{\mu}a_{\nu}\omega_c^2\right),
\end{equation}
where the resonant Rabi frequency $\Omega_c=2|\Omega_{12}|$ with
\begin{equation}
\Omega_{12}=\frac{1}{2}\langle u_1^L|a_{\mu}\partial_{\mu}H({\boldsymbol \lambda}^0)-ia_{\nu}\partial_{\nu}H({\boldsymbol \lambda}^0)|u_2^R\rangle.
\end{equation}
This method can be generalized to quantum systems with more than two energy levels like the Hermitian case \cite{MChen2020}. For instance, to extract the QMT for the lower band of a three-band pH tensor monopole in model \eqref{pHTM}, we need to measure two resonant transitions from the lower energy $E_-$ to the middle (upper) energy $E_0$ ($E_+$).
Here, we shows some numerics for model \eqref{pHHam} based on the above theories. In Fig.~\ref{RF}(a-c), we show the numerical results  of the resonant Rabi frequencies with the corresponding analytic (solid line). Fig.~\ref{RF}(e-d) show the results of the components of the QMT and Berry curvature where the dots and triangles correspond to the numerics and the solid lines correspond to the theoretical predictions.
%%%%%%%%%%%%%%%%%%%%%%%%%%%%%%%%%%%%%%%%%%%%%%%

\end{appendix}

\end{document}